\documentclass[12pt]{iopart}
\usepackage{graphicx}
\usepackage{bm}
\usepackage{iopams}
\usepackage{setstack}

\usepackage{color}

\usepackage{cite}

\begin{document}
\title[Scars in Dirac fermion systems: the influence of an Aharonov--Bohm flux]{Scars in Dirac fermion systems: the influence of an Aharonov--Bohm flux}
\author{Cheng-Zhen Wang$^1$, Liang Huang$^1$,
and Kai Chang{$^2$$^,$$^3$}}
\address{$^1$ School of Physical Science and Technology, and Key
Laboratory for Magnetism and Magnetic Materials of MOE, Lanzhou University,
Lanzhou, Gansu 730000, China}
\address{$^2$ SKLSM, Institute of Semiconductors, Chinese Academy of Sciences, P.O. Box 912, Beijing 100083, China}
\address{$^3$ Synergetic Innovation Center of Quantum Information and Quantum Physics, University of Science and Technology of China,
Hefei, Anhui 230026, China}
\ead{huangl@lzu.edu.cn}
\begin{abstract}
Time-reversal ($\mathcal{T}$-) symmetry is fundamental to many physical processes. Typically, $\mathcal{T}$-breaking for microscopic processes requires the presence of magnetic field. However, for 2D massless Dirac billiards, $\mathcal{T}$-symmetry is broken automatically by the mass confinement, leading to chiral quantum scars. In this paper, we investigate the mechanism of $\mathcal{T}$-breaking by analyzing the local current of the scarring eigenstates and their magnetic response to an Aharonov--Bohm flux. Our results unveil the complete understanding of the subtle $\mathcal{T}$-breaking phenomena from both the semiclassical formula of chiral scars and the microscopic current and spin reflection at the boundaries, leading to a controlling scheme to change the chirality of the relativistic quantum scars. Our findings not only have significant implications on the transport behavior and spin textures of the relativistic pseudoparticles, but also add basic knowledge to relativistic quantum chaos.
\end{abstract}
\pacs{05.45.Mt, 11.30.-j, 71.15.Rf}
\noindent{\it Keywords\/}: relativistic quantum chaos, time-reversal symmetry breaking, chiral scar, semiclassical theory, Aharonov--Bohm phase

\submitto{\NJP}

\maketitle

\section{Introduction}
Time-reversal ($\mathcal{T}$-) symmetry is fundamental and has substantial implications in physical systems
\cite{Griffiths2008,Prigogine1980,Book:quantum,Book:RQM}.
In general, to break the $\mathcal{T}$-symmetry for a microscopic process one needs to involve magnetism \cite{ref:TB}. Without loss of generality we consider a prototype model that is widely used in both classical dynamics and quantum chaos: the billiard system \cite{Sinai:book,Gutzwiller:book,Cvitanovicchaos,Stockmann-book,Haake:book}. For example, a classical picture for a system to break the $\mathcal{T}$-symmetry is a charged particle moving in a magnetic field, whose time-reversed orbit is no longer a solution of the system \cite{RB:1985,BK:1996}. In quantum physics, $\mathcal{T}$-symmetry breaking can be more subtle that the time-reversed trajectory can be the same but the phase of the action integral can be different, such as the Aharonov-Bohm (A-B) effect \cite{Berry1986,Robnik:1992}. The ferromagnetic perturbator in electromagnetic wave analog of Schr\"{o}dinger equation introduces a mechanism to break $\mathcal{T}$-symmetry in microwave billiards \cite{Ott-ferro,Barbara-ferro}.

The discoveries of two-dimensional Dirac fermion systems~\cite{CNN:2011} such as graphene \cite{Novoselovetal:2004,Bergeretal:2004,Novoselovetal:2005,ZTSK:2005,Graphene_review,Graphene_review2,Graphene_review3,Chang:2014}, surface states of 3D topological
insulators~\cite{TIs:RMP,ZhangSC:2011,Chang:2011,Chang:2013}, molybdenum disulfide
(MoS$_2$)~\cite{RRBGK:2011,WKKCS:2012},
HITP [Ni$_3$(HITP)$_2$]~\cite{Sheberlaetal:2014},
and topological Dirac semimetals~\cite{Liuetal1:2014,Liuetal2:2014},
has led to an emerging field of relativistic quantum chaos, where a basic component is the relativistic quantum billiard \cite{Berry:1987,MWZC:2007,PSKY:2008,HLFGA:2009,BDMR:2012,NHLG:2012,XHLG:2013}.
In their seminal work \cite{Berry:1987}, Berry and Mondragon discussed a subtle $\mathcal{T}$-symmetry breaking phenomena, i.e., the 2D massless Dirac particle, when confined within a finite region, automatically breaks the $\mathcal{T}$-symmetry {\em without} the need of involving magnetism. The resulting level spacing statistics of the chaotic Dirac billiard show Gaussian unitary ensemble (GUE) statistics.
Extensive search of this novel $\mathcal{T}$-breaking phenomena in graphene billiards has been carried out \cite{Beenakker-valley,WRAW:2009,HLG:2010,HLG:2011,Rycerz:2012,Rycerz:2013,HXLG:2014}, and found that only in certain cases the valley symmetry can be violated where GUE can be recovered \cite{Rycerz:2012,Rycerz:2013}.

Mathematically, the novel $\mathcal{T}$-symmetry breaking is because the Hamiltonian with the confinement potential, which has to be a scalar 4-potential energy \cite{Berry:1987}, does not commute with the time reversal operator. Consequently, the boundary condition imposed by the confinement potential also does not commute with the time reversal operator. Beside this, Berry and Mondragon provided a semiclassical understanding by considering the phase difference of the plane waves traveling in one direction of the periodic orbit and its time-reversed counterpart \cite{Berry:1987}. They found that for orbits with even number of bounces, the accumulated phase difference between the clockwise and counterclockwise orbit is an integer multiple of $2\pi$, which does not break the time reversal symmetry; only the orbits with odd number of bounces have an additional $\pi$ in the accumulated phase difference, therefore distinguishes the counterclockwise motion from the clockwise motion, and breaks the $\mathcal{T}$-symmetry.
The quantum counterparts of the classical orbits are the quantum scars, which show unusual concentration of the quantum wavefunction on the unstable classical periodic orbits \cite{MK:1979,Heller:1984,MK:1988}.
Following this picture, Xu {\it et al.} investigated the quantum scars in this system, and found an intriguing difference between quantum scars with odd number of reflections at the boundary and those with even reflections, in accordance with the above rationales \cite{XHLG:2013}. These odd-period scars for the Dirac billiard are then named as chiral scars. The chiral property is closely related to the  overall phase change difference of scars. Although the results show distinct difference for the even and odd scars, the $\mathcal{T}$-breaking mechanism from either semiclassical or microscopic perspect is not fully understood.
It has been noted in Ref. \cite{XHLG:2015} that by considering reflection of the planar Dirac spinor wave at the boundary interface of a straight potential jump, there will be a nonvanishing probability current density along the boundary even when the scalar 4-potential energy goes to infinity. Furthermore, the current flow is orientated, i.e., it is fixed to the positive $y$ direction, which is independent to the incident angle that whether it is downward or upward, although the magnitude of the current will be affected. Thus the time-reversed orbit of the planar spinor wave will result in an asymmetric current at the boundary, which breaks the $\mathcal{T}$-symmetry, in accordance of the non-commutable relation between the $\mathcal{T}$-operator and the boundary condition \cite{Berry:1987}.

Here in this paper we revisit this system from both the semiclassical and microscopic aspects to investigate the mechanisms of $\mathcal{T}$-symmetry breaking by scar current analysis and magnetic response, which compensates the rationales of Berry and Mondragon \cite{Berry:1987} with more physical understandings. Furthermore, it provides a controlling scheme which can switch the chiral scars with the non-chiral scars, and also an exact semiclassical formula for the phase accumulation that can be used for level prediction of the relativistic scars, which agrees with the numerical calculations well.
In particular, we consider the chaotic Dirac A-B billiard with a vanishing inner radius. Therefore, we introduce an additional phase caused by the magnetic flux, and in the mean time the orbits, thus the scars, are not perturbed. An experimentally feasible setup would require a finite inner radius. However, insofar as the inner region for the magnetic flux does not intersect with the orbit of the scar, it has little influence to the scar.

\section{Model and methods}
To be concrete, the chaotic Dirac A-B billiard is as follows.
The system consists of a single massless spin-half particle with charge $q$ confined by hard walls (infinite mass confinement) in a heart-shaped or Africa domain ($w$ plane) whose classical dynamics is chaotic, and threaded by a single line of magnetic flux $\Phi$ at the origin. The position of the line of magnetic flux is a singular point. Therefore, we exclude this point by considering an inner disk of infinite mass potential with a vanishing inner radius centered at this point. Thus the flux can introduce a modulating phase, in the meantime, as it is just a single point on the 2D billiard, it does not exert much spatial perturbations to the scarring states. The billiards in the $w=u+iv$ plane can be conformally transformed from a unit disk on the complex $z=x+iy$ plane,
\begin{eqnarray}\label{CTM-exp}
w(z) = \frac{z+bz^2+ce^{i\delta}z^3}{\sqrt{1+2b^2+3c^2}},
\end{eqnarray}
where for the heart-shaped billiard $b=0.49$, $c=\delta=0$, and for the Africa billiard $b=c=0.2$, $\delta=\pi/3$. Please note that with the above parameters these two billiards have chaotic classical dynamics \cite{Robnik83,Bruus-heartchaotic}.

For the magnetic flux, we choose a non-divergent gauge in which the lines of the vector potential $\bi{A}$ are the contours of a scalar function $F(u,v)$:
\begin{eqnarray}
\bi A({u,v}) = \frac{\Phi}{2\pi} (\frac{\partial F}{\partial v}, \frac{-\partial F}{\partial u}),
\end{eqnarray}
and $F$ satisfies that $\nabla^{2}_{uv}F=-2\pi \delta(u)\delta(v)$ \cite{Berry1986}.

The Hamiltonian for the confined Dirac particle is
\begin{eqnarray}
\hat{H} = v_F\hat{\bi{\sigma}}\cdot (\hat{\bi{p}} - \frac{q}{c}\bi{A}) + V(u,v)\hat{\sigma}_z,
\end{eqnarray}
where $v_F$ is Fermi velocity,
$\hat{\bi{\sigma}}=(\hat{\sigma}_x, \hat{\sigma}_y)$ and $\hat{\sigma}_z$ are the Pauli matrices, $V(u,v)=0$ within the billiard, and $V(u,v)=\infty$ outside the confinement region.
The Dirac equation in the billiard can be written as
\begin{equation}\label{eq:DiracAB}
v_F\hat{\bi{\sigma}}\cdot (\hat{\bi{p}} - \frac{q}{c}\bi{A}) \Psi = E\Psi,
\end{equation}
where $\Psi=[\Psi_1,\Psi_2]^T$ is the spinor wavefunction, and the boundary condition is \cite{Berry:1987}
\begin{equation} \label{Heart-boundcond}
\frac{\Psi_2}{\Psi_1}|_{\partial D} = ie^{i\widetilde{\theta}(s)},
\end{equation}
where $s$ is the coordinate that describes the arc length of the boundary, starting from the cross point of the boundary with positive $u$-axis; $\widetilde{\theta}(s)$ is the angle to the positive $u$-axis for the normal vector at $s$.
When being acted upon by the Hamilton operator $\hat{H}$ again, Equation (\ref{eq:DiracAB}) becomes
\begin{eqnarray*}
\nabla^{2}_{uv}\Psi(w) - 2i\alpha \big(\frac{\partial F}{\partial v}\frac{\partial}{\partial u} - \frac{\partial F}{\partial u}\frac{\partial}{\partial v}\big)\Psi(w) + i\alpha\hat{\sigma}_x\hat{\sigma}_y\nabla^{2}_{uv}F - \\ \alpha^{2}\big[\big(\frac{\partial F}{\partial u}\big)^{2} + \big(\frac{\partial F}{\partial v}\big)^{2}\big]\Psi(w) + k^2\Psi(w) = 0.
\end{eqnarray*}
where $\alpha={q\Phi}/(h c)$ and $k = {E}/(\hbar v_F)$. Note that the term $i\alpha\hat{\sigma}_x\hat{\sigma}_y\nabla^{2}_{uv}F$ is particular to the Dirac A-B billiard, which is not present in the Shr\"{o}dinger A-B billiard \cite{Berry1986}. However, since $\nabla^{2}_{uv}F=-2\pi \delta(u)\delta(v)$, it is singular at the origin and is zero otherwise. Practically, by setting a inner disk with radius $\xi \ll 1$ of infinite mass potential, the billiard region that we are interested excludes this singular point. Note that the inclusion of the A-B flux can have two different types of boundary conditions around the singular point. Except introducing an infinite mass boundary for the inner disk and letting the radius go to zero, which is relevant for our case where the A-B flux only contributes to a global phase, there is a different setup for the boundary condition of the A-B flux in the quantum field theory where further interactions need to be considered to calculate the vacuum energy \cite{LR:1998,BFKS:2000}. Then in the $\xi \rightarrow 0$ limit, it has little perturbations to the wavefunctions. Therefore, in the following treatment to solve the eigenvalue and eigenfunctions, this term has been omitted.

Changing back to the disc region in the $z$-plane $\bi{r}=(x,y)$ is a straightforward procedure based on $w(z)$. We obtain
\begin{eqnarray*}
\nabla^{2}\Psi(\bi{r}) - 2i\alpha\big(\frac{\partial F}{\partial y}\frac{\partial}{\partial x} - \frac{\partial F}{\partial x}\frac{\partial}{\partial y}\big)\Psi(\bi{r}) - \\ \alpha^{2}\big[\big(\frac{\partial F}{\partial x}\big)^{2} + \big(\frac{\partial F}{\partial y}\big)^{2}\big]\Psi(\bi{r})
 + k^2|w^{'}(z)|^{2}\Psi(\bi{r}) = 0,
\end{eqnarray*}
where the last term includes the nonuniform part $|w^{'}(z)|^{2}$ originated from the chaotic boundary in the $w$ plane. In particular, $F$ can be chosen as $F(\bi{r})=-\ln|\bi{r}|$ in the $z$ plane, so in polar coordinates the above equation can be written as
\begin{eqnarray}\label{eq:chaoticAB}
\nabla^{2}\Psi(r,\theta) - \frac{2i\alpha}{r^2}\frac{\partial \Psi(r,\theta)}{\partial \theta} - \frac{\alpha^{2}}{r^2}\Psi(r,\theta) + k^2|w^{'}(z)|^{2}\Psi(r,\theta) = 0.
\end{eqnarray}
To solve the above equation, we expand $\Psi$ in terms of eigenfunctions $\psi_{lm}(r,\theta)$ of the circular Dirac A-B billiard of the unit disc with a vanishing inner radius (\ref{Appendix-A}), whose corresponding eigenvalues are $\mu_{lm}$, with $l$ and $m$ relevant quantum numbers. We have
\begin{eqnarray}\label{eq:superposition}
\Psi(r,\theta) = \sum^{\infty}_{l=-\infty}\sum^{\infty}_{m=1} c_{lm}\psi_{lm}(r,\theta),
\end{eqnarray}
where $c_{lm}$ are the expansion coefficients. Substituting Equation (\ref{eq:superposition}) into Equation (\ref{eq:chaoticAB}) leads to
\begin{eqnarray}\label{eq:eig0}
\frac{\nu_{lm}}{k^2} - \sum_{l^{'}m^{'}}M_{lml'm'}\nu_{l'm'} = 0,
\end{eqnarray}
where $\nu_{lm} = \mu_{lm}c_{lm}$, and
\begin{eqnarray}\label{eq:Melement}
M_{lml'm'} =  \frac{N_{lm}N_{l'm'}}{\mu_{lm}\mu_{l'm'}}\int^{1}_{0}rdr\int^{2\pi}_{0}d\theta|w'(z)|^2\cdot e^{i(l'-l)\theta}\nonumber\\
\cdot\Big\{\big[J_{\nu}(\mu_{lm}r) + \beta_{lm} N_{\nu}(\mu_{lm}r)\big]
\cdot \big[J_{\nu'}(\mu_{l'm'}r)+\beta_{l'm'}N_{\nu'}(\mu_{l'm'}r)\big]
\nonumber\\ +\big[J_{\nu+1}(\mu_{lm}r)+\beta_{lm}N_{\nu+1}(\mu_{lm}r)\big]
\cdot\big[J_{\nu'+1}(\mu_{l'm'}r)+\beta_{l'm'}N_{\nu'+1}(\mu_{l'm'}r)\big]\Big\}.
\end{eqnarray}
The angular integration in Equation (\ref{eq:Melement}) can be calculated analytically, which yields
\begin{eqnarray}\label{Angle-integral}
I=\int^{2\pi}_{0}d\theta|w'(z)|^2\cdot e^{i(l'-l)\theta} 
=
\left \{ \begin{array}{ll}
2\pi(1+4b^2r^2+9c^2r^4) \quad &l = l',\\
2\pi(2br + 6bcr^3 e^{\pm i\delta}) \quad &l=l'\pm1,\\
2\pi(3cr^2e^{\pm i\delta}) \quad & l=l'\pm2.
\end{array} \right.
\end{eqnarray}\\
Substituting $I$ into (\ref{eq:Melement}) and integrating over variable $r$ (we use the simplified form of radical function in \ref{Appendix-A} instead of that in Equation (\ref{eq:Melement})), we can obtain the $M$ matrix. Equation (\ref{eq:eig0}) can be written in the form of eigen-equation: $M V_n=\lambda_n V_n$, where $k_n=1/\sqrt{\lambda_n}$, $c_{n,lm}=V_{n,lm}/\mu_{lm}$. Correspondingly, we can get the eigen-energy as $E_n=\hbar v_F k_n$ of the original chaotic Dirac A-B billiard, and the eigen-state in the $w$ plane can be obtained from that in the $z$ plane: $\Psi_n(u,v)=\Psi_n(x(u,v),y(u,v))$, and
$\Psi_n(r,\theta)=\Sigma_{lm} c_{n,lm}\psi_{lm}(r,\theta)$.
\section{Results}
Once the eigenstates are obtained, we plot each of them and identify those localized on classical orbits---the scarring states. As proposed in Ref. \cite{XHLG:2013}, we use $\eta$ to characterize the wavevector difference between the repetitive scars on the same orbits, which is defined as
\begin{eqnarray}\label{eq:eta}
\eta &= \frac{|k_n - k_0|}{\delta k}-\Big[\frac{|k_n - k_0|}{\delta k}\Big],
\end{eqnarray}
where [$x$] denotes the largest integer less than $x$, $k_0$ is the wavevector for a scar setting as the reference point, $k_n$ is the wavevector for repetitive scars on the same orbit, $\delta k = 2\pi/L$ and $L$ is the orbital length. Typically, $\eta$ has the values of either close to 0 or 1. However, for scars on odd orbits (chiral scars) the feature is that $\eta$ can take values around 0.5 \cite{XHLG:2013}.
This $0.5$ value of $\eta$ has been argued as due to the time-reversal symmetry breaking of the scars on odd orbits \cite{XHLG:2013}, which semiclassically has been proposed by Berry and Mondragon \cite{Berry:1987}, that the spinor plane waves with odd number of bounces have an additional $\pi$ in the phase difference between counterclockwise and clockwise orbits while the plane waves with even number of bounces have not. 
Note that the phase change here is caused by the boundary-spin interaction at the boundary. During each collision, the phase difference between the counterclockwise reflection and its time reversed counterpart has an additional $\pi$ contribution. This phase $pi$ leads to the spin polarization at the boundary. Also, we can see that for a scar on an orbit with even number of reflections, the spin-boundary interaction contributes to an integer multiple of $2\pi$ for the phase difference of the counterclockwise orbit and its clockwise counterpart. Thus for these orbits, the time-reversal symmetry is preserved. However, for the scars with odd number of reflections, the boundary phases contribute an additional $\pi$, leading to the $\mathcal{T}$-symmetry breaking and also a chiral signature of the scar (Details about the local and global phase changes are discussed in \ref{Appendix-B}).

\subsection{Current analysis of scars}
To investigate the phase of the scarring eigenstates, we examine their local current flows. The current operator is given by
\begin{eqnarray}\label{eq:current-operator}
\hat{\bi{u}} = \bi{\nabla_p} \hat{H} = v_F \hat{\bi{\sigma}},
\end{eqnarray}
and the local current for state $\Psi(\bi{w})$ can be defined as the expectation value of $\hat{\bi{u}}$~\cite{Berry:1987}:
\begin{eqnarray}\label{eq:local-current}
\bi{u}
&\equiv v_F(\psi_{1}^{\ast}(\bi{w}), \psi_{2}^{\ast}(\bi{w})) \hat{\bi{\sigma}} \left(\begin{array}{c}
\psi_{1}(\bi{w})\\
\psi_{2}(\bi{w})
\end{array} \right)\nonumber\\
&= 2v_F[\Re(\psi_{1}^{\ast}(\bi{w}) \psi_{2}(\bi{w})), \Im(\psi_{1}^{\ast}(\bi{w}) \psi_{2}(\bi{w}))].
\end{eqnarray}

\begin{figure}
	\centering
  \includegraphics[width=\textwidth]{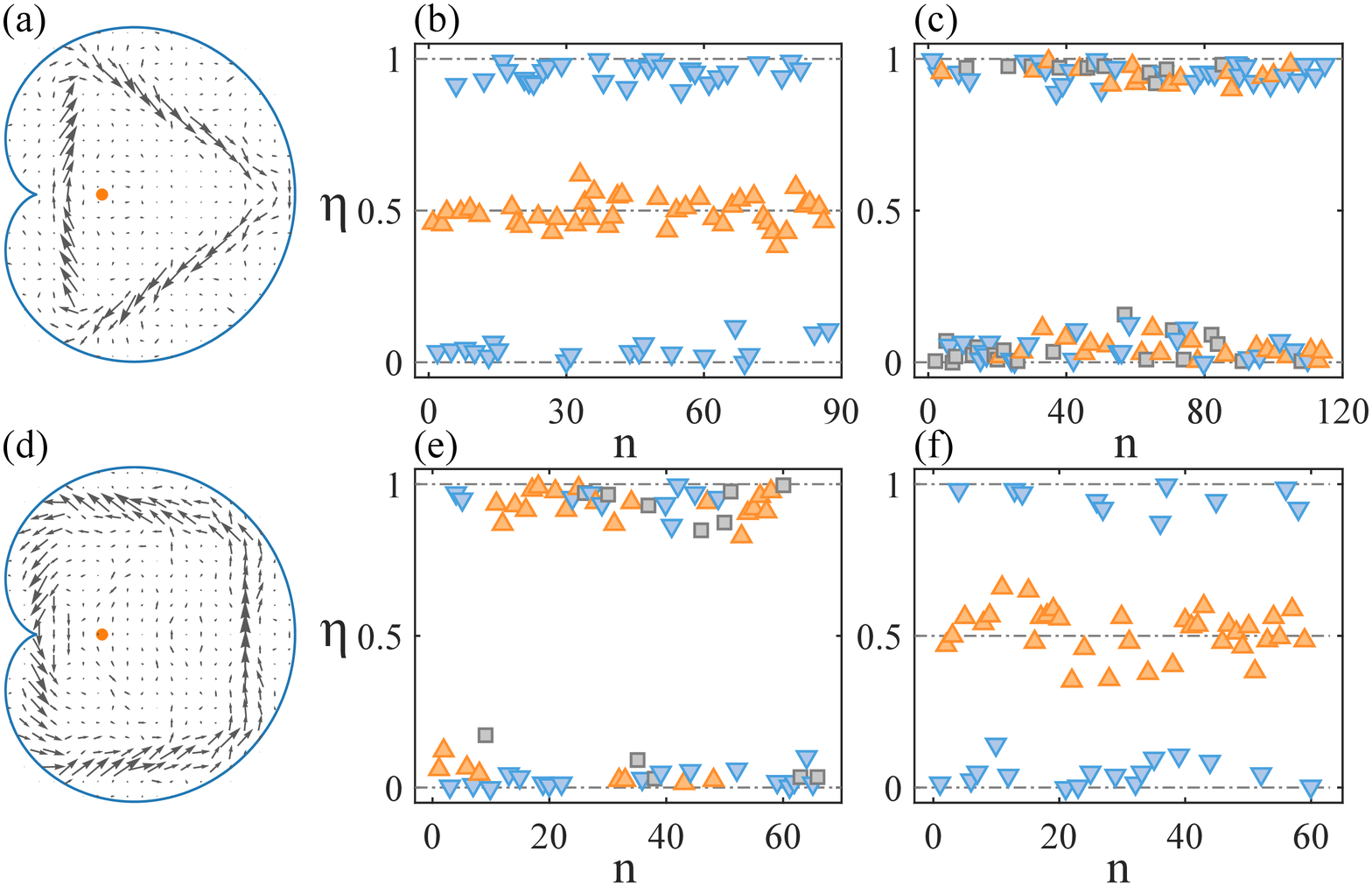}
  \caption{The current of scars (a,d), and the corresponding $\eta$ values at $\alpha = 0$ (b,e) and $\alpha = 1/4$ (c,f). The first row is for a period-3 scar, and the second row is for a period-4-II scar. The orange up-triangles are for scars with counterclockwise flow, the blue down-triangles are for scars with clockwise flow, and the gray squares are for scars whose current orientation is hard to distinguish. The reference state is chosen (arbitrarily) from the scars with clockwise flow. The orange dot in (a), (d) is the origin with single magnetic flux.}
	\label{fig:ABscar}
\end{figure}

A systematic investigation of the local current flow for scarred states indicates that the current of most scars has a definitive orientation, either clockwise or counterclockwise, as shown in figure~{\ref{fig:ABscar}} (a) and (d) for period-3 orbit and period-4-II orbit, respectively.
We estimated the relation between scar wavevector difference $\eta$ and the scar orientation defined by its current flow. In figure~{\ref{fig:ABscar}} the scarring states with counterclockwise flow are marked as orange up triangles and those with clockwise flow are marked as blue down triangles. It is found that for even bounce scars, the wavevector difference $\eta$ is always $0$ or $1$, regardless of relative current orientation [figure~{\ref{fig:ABscar}} (e)]; while for odd bounce orbit, when two scars have the same current orientation, $\eta = 0$ or $1$, while if two scars have opposite current orientation, then $\eta = 1/2$, as shown in figure~{\ref{fig:ABscar}} (b), indicating $\mathcal{T}$-symmetry breaking from the semiclassical point of view. This current orientation analysis confirms that $\eta = 1/2$ is resulted from the $\pi$ phase difference of the opposite current orientation of odd bounce scars.

\subsection{Scar chirality change by magnetic flux}

A natural question is that can this phase be compensated by the magnetic flux? In particular, we consider a magnetic flux $\alpha$ (in units of magnetic flux quanta $\phi_0 \equiv hc/q$) and a winding number $W$ of a certain orbit around this flux, the phase gain caused by the magnetic flux is $2\pi W\alpha$. For a time reversed orbit, $W$ changes sign, thus the phase  difference between these two orbits with opposite orientation is $4\pi W\alpha$. Therefore, for the case of $W=1$, if $\alpha=1/4$, then it will introduce a $\pi$ phase difference. If the phase exerted by boundary-spin interaction in spin is equivalent to that caused by the magnetic flux, then in the case of $W=1$ and $\alpha=1/4$, the odd orbit scars will lose its chiral character, while the even orbit scars will become chiral.

As shown in figure \ref{fig:ABscar}, when there is no magnetic flux, $\eta$ attains 0.5 value for the period-3 scar, indicating the chirality of this scar. However, when $\alpha=1/4$, the data points of $\eta \sim 0.5$ have been disappeared, leading to
a superficial time-reversal preservation. While for the period-4-II scar, the data points of $\eta \sim 0.5$ do not present for $\alpha=0$ but emerge for $\alpha=1/4$. This indicates that although originated from different mechanism, the boundary-spin interaction induced phase is equivalent to that of magnetic flux.
It is noticed that for scars without chiral nature, the two flow orientations are mixed. While
for scars with a chiral nature, i.e., period-3 scars with $\alpha=0$ and period-4-II scars with $\alpha=1/4$, the scars with different orientation are well separated. One set of the scars attains a $0.5$ value for $\eta$, while the other set attains values of $0$ or $1$.

\begin{figure}
\centering
  \includegraphics[width=\textwidth]{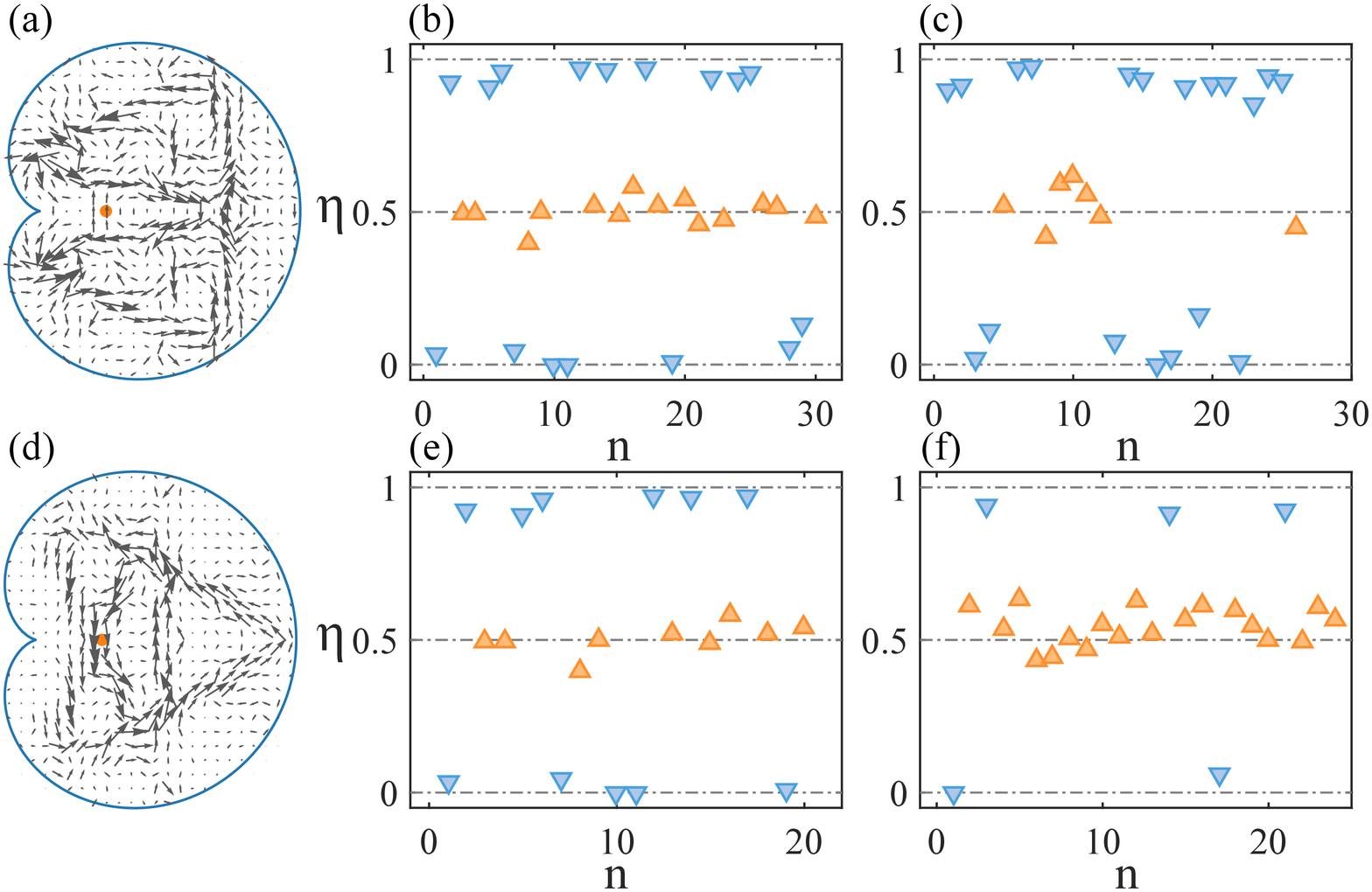}
\caption{The same plots as in figure \ref{fig:ABscar} except that the first row is for the period-5-I scar, and the second row is for the period-5-II scar. }
  \label{fig:ABscar5}
\end{figure}
Figure \ref{fig:ABscar5} plots the same quantities as in figure \ref{fig:ABscar} but for two period-5 scars. Surprisingly, $\eta$ for $\alpha=0$ and $\alpha=1/4$ appear the same. A more detailed examination reveals that, for the period-5-I orbit, the flux is outside and not circulated by the orbit, therefore the flux has no effect to this scar. However, for the period-5-II orbit, it circulates the flux twice, i.e., $W=2$, thus when $\alpha=1/4$ the phase difference between the counterclockwise orbit and the clockwise orbit is $4\pi W\alpha = 2\pi$, which does not change the chirality of the scars.

\subsection{Semiclassical theory of scars}

\begin{figure}
\centering
  \includegraphics[width=\textwidth]{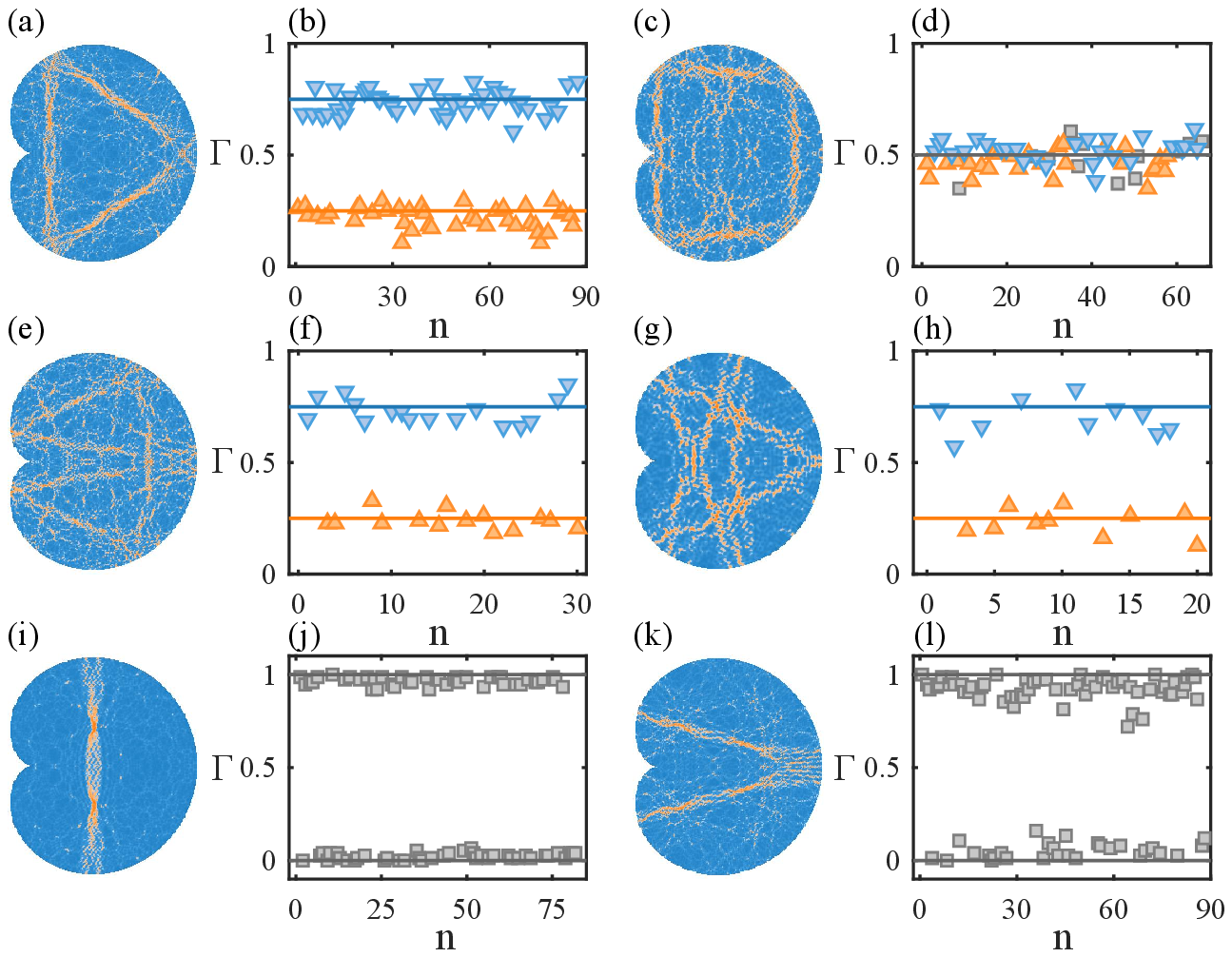}
\caption{$\Gamma$ values for different scars, (a) period-3, (c) period-4-II, (e) period-5-I, (g) period-5-II, (i) period-2, (k) period-4-I. Orange up-triangles and blue down-triangles represent scars with counterclockwise and clockwise current orientation, respectively. Gray squares represent scars without obvious current orientation.
The horizontal solid lines indicate the semiclassical predictions in Table~{\ref{tab:scar-level-heart}}.}
  \label{fig:k_Maslov}
\end{figure}

Phenomenologically, as the phase caused by the boundary-spin interaction is equivalent to that by the magnetic flux, we can include it in the phase shift formulae \cite{Stockmann-book,Borondo2005,Borondo2006,Richter-tracespin},
\begin{eqnarray}\label{eq:action}
\Delta \Phi = \frac{1}{\hbar}S - \frac{\sigma \pi}{2} +2\pi\beta 
= k\cdot L +2\pi W\alpha - \frac{\sigma \pi}{2} +2\pi\beta,
\end{eqnarray}
where the action $S = \oint\bi{p}\cdot d\bi{q}=
\hbar\oint\bi{k}\cdot d\bi{q} + \frac{q}{c}\oint\bi{A}\cdot d\bi{q}$ \cite{Berry1986}, $W$ is the winding number encloses the flux, $\sigma$ is the Maslov index that related to the conjugate points along the orbit and is canonical invariant \cite{Littlejohn}. Here in the heart-shaped billiard, $\sigma$ equals to the number of reflections along the complete orbit \cite{Bruus1996}. The infinite mass (or hard wall) reflection only contributes phase in the spin term, thus has no contribution to the Maslov index, and $2\pi\beta$ represents the phase accumulation of spin reflection at the boundary, whose value depends on the particular orbit and current orientation. Note that because of the chiral effect caused by spin boundary interaction, there is a $\pi$ difference in the term $2\pi\beta$ between the reversed odd orbits (\ref{Appendix-B}). For semiclassically allowed orbits the phase accumulation around one cycle should be multiple integers of $2\pi$, i.e., $\Delta\Phi = 2\pi n$, $n=1,2,\cdots$ to ensure that the wavefunction is single-valued. Thus
\begin{eqnarray}\label{eq:action1}
k &= \frac{2\pi}{L}(n - W\alpha +\frac{\sigma}{4} - \beta).
\end{eqnarray}
In the case of zero magnetic flux ($\alpha=0$), we define $\Gamma = mod(kL/2\pi,1)=mod(\sigma/4 - \beta,1)$, which relates the semiclassical quantity $\sigma$ (the number of conjugate point on the orbit) and $\beta$ from the relativistic quantum dynamics. Here we list the values of parameters $\sigma$, $\beta$ and $\Gamma$ (via $mod(\sigma/4 - \beta,1)$) in Table~{\ref{tab:scar-level-heart}} for different orbits. Alternatively, the values of $\Gamma$ can be obtained numerically through $mod(kL/(2\pi),1)$ from the eigenwavevectors of the corresponding scars. The results are shown in figure \ref{fig:k_Maslov}. We can see that the $\Gamma$ values obtained from numerical calculations agree with the semiclassical theory well.
\begin{table}
\caption{\label{tab:scar-level-heart}The values of $\sigma$, $\beta$ and $\Gamma$ are for different orbits (shown in figure \ref{fig:k_Maslov}). (+,-) denote counterclockwise and clockwise orientation, respectively.}
\renewcommand{\arraystretch}{1.2}
\begin{indented}
\lineup
\item[]\begin{tabular}{@{}ccccccc}
\br
Orbits &2 &3 &4-I &4-II &5-I &5-II\\
\mr
$\sigma$ &2 &3 &4 &4 &5 &5\\

$\beta$  &1/2 & \begin{tabular}{cc}{1/2 (+)}\\ {0 (-)}\end{tabular} &1 &1/2 &\begin{tabular}{cc}{1 (+)}\\{1/2 (-)} \end{tabular} &\begin{tabular}{cc}{1 (+)}\\ {1/2 (-)} \end{tabular}\\
$\Gamma$ &{0, 1} & \begin{tabular}{cc}{1/4 (+)}\\{3/4 (-)}\end{tabular} &0, 1 &\begin{tabular}{cc}{1/2 (+)}\\{1/2 (-)}\end{tabular} &\begin{tabular}{cc}{1/4 (+)}\\{3/4 (-)}\end{tabular} &\begin{tabular}{cc}{1/4 (+)}\\{3/4 (-)}\end{tabular} \\
\br
\end{tabular}
\end{indented}
\end{table}

\subsection{Magnetic control of scars}

Now we examine the wavevector changes of scars tuned by a magnetic flux at the origin. The wavevector difference of reversed scars of the same type is denoted as
\begin{equation}\label{eq:k-alpha-reverse}
  \Delta k = \cases{
  2\pi (\Delta n - 2W\alpha)/L &\text{even bounces},\cr
  2\pi (\Delta n - 2W\alpha + \Delta \beta)/L \quad &\text{odd bounces}.}
\end{equation}
where $n$ is an integer, and $\Delta \beta = {1}/{2}$ for odd orbits.
Thus whenever $|2W\alpha|=1/2$ for an orbit, the corresponding scars will interchange between chiral and non-chiral characters, as demonstrated in figure \ref{fig:ABscar}.
\begin{figure}[!htp]
\centering
  \includegraphics[width=0.8\textwidth]{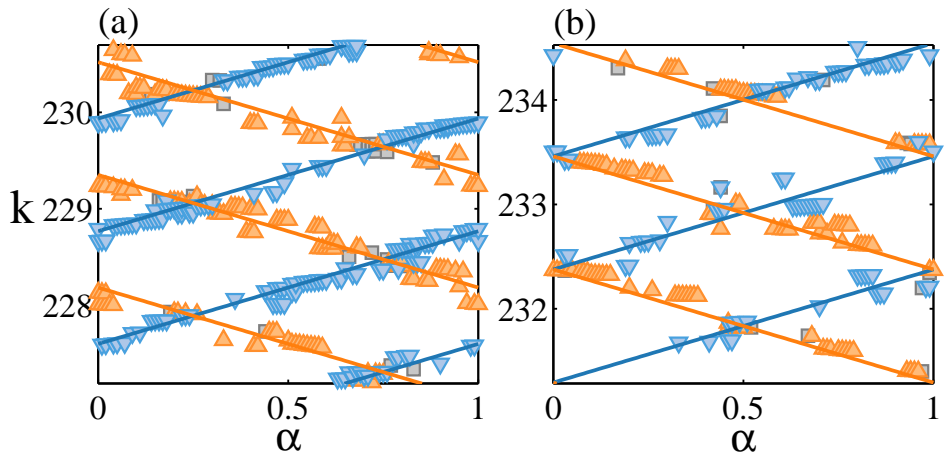}
\caption{The relation between wavevector $k$ and the magnetic flux $\alpha$, for (a) the period-3 scar shown in figure \ref{fig:ABscar}(a), and (b) the period-4-II scar shown in figure \ref{fig:ABscar}(d). The orange up-triangles indicate scars with counterclockwise flow, where $W=1$, and blue down-triangles are the scars with clockwise flow, where $W=-1$. The gray squares are the scars that is difficult to identify the flow orientations. The solid lines are theoretical predictions of Equation (\ref{eq:k-alpha}). The step in the variation of $\alpha$ is 0.01.}
  \label{fig:k-alpha}
\end{figure}

From Equation (\ref{eq:action1}), for a scar with wavevector $k_0$ at $\alpha=0$, as the magnetic flux $\alpha$ is increased, the same scar would appear if the wavevector approximately follows
\begin{eqnarray}\label{eq:k-alpha}
k = k_0 - W \alpha \frac{2\pi}{L},
\end{eqnarray}
as $\beta$ depends only on the orbit and is fixed to a particular value for a given orbit. The system is periodic for magnetic flux varying from 0 to 1.
We have varied the magnetic flux systematically, and for each case, identified the scar on the same orbit in a certain wavevector (energy) range and identified their flow orientation. The corresponding wavevector and magnetic flux for the same type period-3 and period-4-II scars [figure \ref{fig:ABscar}(a) (d)] are plotted in figure \ref{fig:k-alpha}. The solid lines are from Equation (\ref{eq:k-alpha}). One can see that the numerics follow the theory well. Note that Equation (\ref{eq:k-alpha}) holds for both odd periodic and even periodic orbits. The difference, however, comes from the initial $k_0$ value. From figure \ref{fig:k-alpha} it is clear that for the scars on any orbit, there are actually two sets of scars, one with counterclockwise flow, i.e., $W=1$, where $k$ decreases linearly with increasing $\alpha$; the other with clockwise flow that $W=-1$, where $k$ increases with increasing $\alpha$. For each set, if one fixes the magnetic flux and examines the eigenstates, the scar repeats itself when $\Delta k = 2\pi/L$ approximately holds. However, when there is no magnetic flux, the two sets of odd periodic scars intersect each other, leading to $\Delta k = \pi/L$ if the flow orientation is not distinguished. But if we regard the two sets are different scars, then for each set, we recover $\Delta k = 2\pi/L$. For the even period scars, the two sets appear parallel to each other, i.e., they may appear at the same set of $k_0$ values with $2\pi/L$ intervals, although at each $k_0$, typically only one scar can be found.

The wavevector $k$ for the scar goes down as $\alpha$ increases for $W=1$, while it goes up for $W=-1$. Therefore, the two lines cross each other at certain points. For the period-3 scar, the cross points are $\alpha=0.25$ (corresponding to a $\pi$ phase difference) and $\alpha=0.75$. It is noted that at the cross point, for some of the scars it is difficult to identify the flow orientation. While for the period-4-II scar, the cross points are at $\alpha=0$ and $\alpha=0.5$. For the period-3 scar, if $\alpha$ is shifted by $0.25$, then the $k$-$\alpha$ relation will behave similarly to that for the period-4-II scar. Thus the behavior of period-3 scars at $\alpha=0.25$ is similar to that of the period-4-II scars at $\alpha=0$, and vice versa. In this sense, the magnetic flux interchanges the chiral and nonchiral nature of the period-3 scar and the period-4-II scar by exerting a flux of $\alpha=0.25$.
Now the effect of the boundary induced phase $\beta$ is quite clear, e.g., compared to the period-4-II scar, it shifts the overall pattern of the period-3 scar leftwards from $\alpha=1/4$ to $\alpha=0$, with all other features kept except $k_0$ and $L$ taking different values.

\begin{figure}[h]
\centering
  \includegraphics[width=0.8\textwidth]{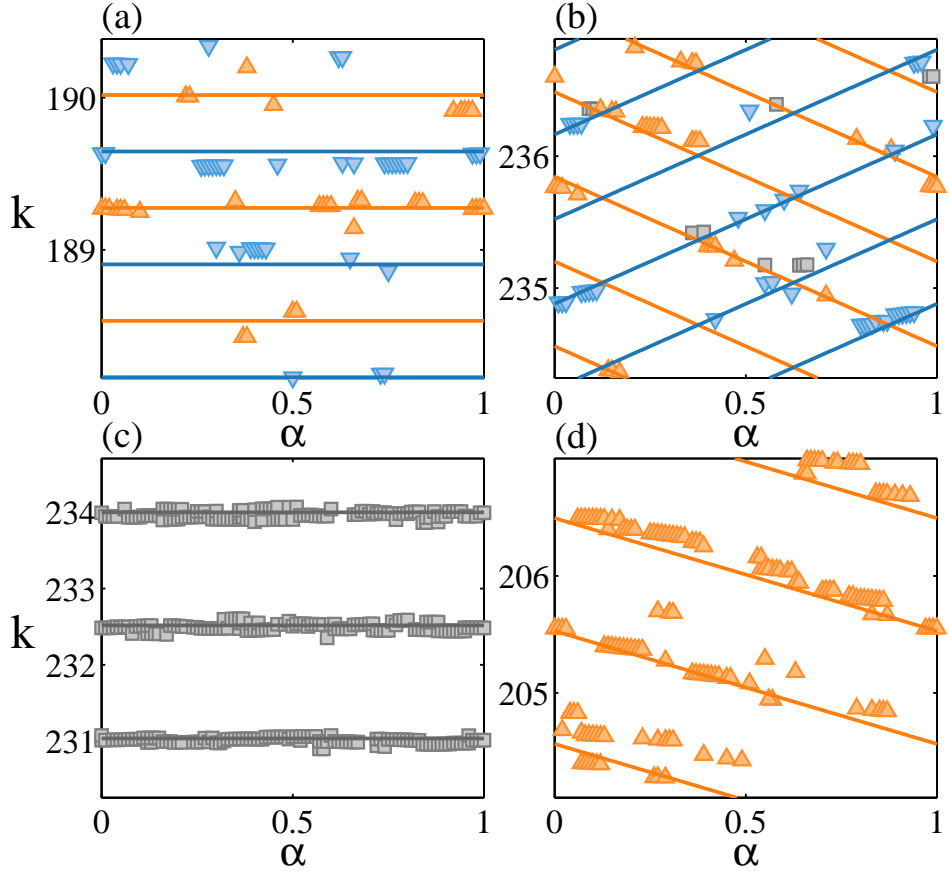}
\caption{The same plots as figure \ref{fig:k-alpha}, for (a) the period-5-I scar shown in figure \ref{fig:ABscar5}(a) with $W=0$, (b) the period-5-II scar shown in figure \ref{fig:ABscar5}(d) with $W=\pm 2$, (c) a period-2 orbit, and (d) the edge state with $W=1$. }
  \label{fig:k-alpha5}
\end{figure}

Figure \ref{fig:k-alpha5} shows the $k$-$\alpha$ relation for another four typical states: the period-5-I scar, the period-5-II scar, a period-2 bouncing ball scar, and an edge state. Since the period-5-I scar [figure \ref{fig:k-alpha5}(a)] and the period-2 bouncing ball scar [figure \ref{fig:k-alpha5}(c)] do not circulate the flux, e.g., $W=0$, thus $k$ does not change with $\alpha$, which agrees with the data. For the period-5-I scar, the state with counterclockwise flow and that with clockwise flow succeeds to each other, i.e., one row with counterclockwise flow (orange up-triangle), then next row with clockwise flow (blue down-triangle) at an wavevector interval $\Delta k = \pi/L$, and vice versa. For the period-2 bouncing ball scar, since there are no specific orientation of the flow, they are represented by gray squares and the wavevector difference between the neighboring rows is $\Delta k = 2\pi/L$. For the period-5-II scar [figure \ref{fig:k-alpha5}(b)], as $W=\pm 2$, the slope is larger, and the cross points are at $\alpha=1/8,3/8, 5/8, 7/8$, i.e, four cross points instead of two for the $W=\pm 1$ cases. Therefore, for the period-5-II scar, it will lose chirality at $\alpha=1/8$ rather than $\alpha=1/4$ for the period-3 scars. For the edge state [figure \ref{fig:k-alpha5}(d)], since it always has a counterclockwise flow at the boundary, the time-reversed state is no longer a solution of the system. Therefore, $W$ can only take the value of $1$, and consequently, in the figure of $k$-$\alpha$ relation, there is only one set of the lines that $k$ decreases with $\alpha$ and the wavevector difference of neighboring lines is about $\Delta k = 2\pi/L$.

Similar results are also obtained in the Africa billiard which has no reflection symmetry (\ref{Appendix-C}).

\section{Experimental realization}
Experimentally, such a novel $\mathcal{T}$-breaking effect can be investigated using topological insulators (TI). In particular, consider a 2D surface supporting the edge states of a 3D topological insulator, whose quasiparticles can be described by the 2D massless Dirac equation (with a 90 degrees rotation of the spins). The mass confinement can be realized by depositing a ferromagnet insulator cap
layer on top of the TI outside the billiard (or quantum dot) region \cite{XNLZ:2012,KRRZ:2013,FL:2013}, where the exchange coupling $V\hat{\sigma}_z$ induced by the ferromagnet insulator can serve as the mass confinement. Although for simplicity the theoretical treatment requires the mass potential goes to infinity, in realistic cases, as far as the energy of the concerned states is much smaller than the gap, the phenomenon would be basically the same. For applying the magnetic flux, in general, the area of the flux threading the surface can be finite, insofar as it is not on the orbit of the scar. For typical scars such as the period-3 and period-4-II scars shown in figure \ref{fig:ABscar}, as they have a large interior, they are less likely to be affected by opening a hole in the middle to exert the magnetic flux.

\section{Discussions and conclusion}
Through extensive computations and physical analysis of the chaotic Dirac A-B billiard, the whole picture of the mechanism of $\mathcal{T}$-symmetry breaking emerges. To be specific, for the Dirac billiard confined by the infinite scalar 4-potential, or mass potential, the Hamiltonian does not commute with the $\mathcal{T}$-operator, as the confinement mass potential will acquire a sign change after the $\mathcal{T}$-operation, which can be corroborated by fact that the boundary condition derived from the mass potential confinement does not commute with the $\mathcal{T}$-operator too. From the local physical interaction point of view, each reflection at the boundary breaks the time-reversal symmetry as it contributes to an oriented flow at the boundary whose direction is independent of the incident angle.
Furthermore, as the spin of a free Dirac particle is polarized along its momentum, the reflection at the boundary induces the boundary-spin interaction, thus each reflection is accompanied with an additional phase $\phi$ in the action integral of the particle. The reversed orbit will acquire another phase $\widetilde{\phi}$ at this point. The phase difference between the counterclockwise reflection and its time reversed reflection at the same boundary point has a $\pi$ contribution. Therefore, for a scar on an orbit with even number of reflections, the total effect of these phases contributes to an integer multiple of $2\pi$ for the phase difference of the counterclockwise orbit and its clockwise counterpart. Thus for these orbits, the time-reversal symmetry is preserved. However, for the scars with odd number of reflections, the boundary phases contribute an additional $\pi$, leading to the $\mathcal{T}$-symmetry breaking and also a chiral signature of the scar.
A natural question is that can this boundary-spin interaction induced phase be compensated by a magnetic flux? The answer is yes. As we have demonstrated, the $\pi$ phase difference between the counterclockwise and clockwise orbits with odd number of reflections can be annihilated completely by a properly added magnetic flux, i.e., the chiral scar loses its chirality, while the non-chiral scars can attain the chirality under certain cases. However, depending on the location of the flux threading the billiard, the winding number for an orbit around this flux can be highly nontrivial. As we show, for a given A-B billiard, the winding numbers can be zero, one, two, and so on, which has significant implications in their response to the flux. The underling rationale is that, phenomenologically, the boundary induced phase can be included into the action integral. Insofar as it is in the action integral, it loses the complexity when generating it, and is equivalent to the phase terms caused by the path integral of the momentum, and thus to the phase from the magnetic flux.
Note that besides the scars on the periodic orbits, there is another class of states, edge states, that always have a counterclockwise flow localized at the boundary, which breaks the time-reversal symmetry as their time-reversed states are no longer solutions for the system. These states have nonzero wavefunctions at the boundary, in contrast to zero wavefunctions at the boundary for the Shr\"{o}dinger billiard with infinite confinement potential.

For the Dirac billiard system, the chirality is fundamentally related with the time-reversal symmetry. The time-reversal operator changes the sign of the confinement potential $V$ and the direction of local flow for the scarring states. The parity operation is effectively the combination of time-reversal operation and mirror reflection. From the semiclassical point of view, for a particular scar, if the billiard has a reflection symmetry, e.g., the heart-shaped billiard, since the mirror reflection becomes identical operation, then the parity operation becomes equivalent to the time-reversal operation. Thus if the system or the state is invariant under the parity operation, it will also be invariant under time reversal operation, such as for the even period scars that at a given energy level the flow orientation can be either clockwise or counterclockwise. For odd period scars, both the parity symmetry and the time-reversal symmetry are broken, arousing a chiral signature for these scars and at a given energy level only one orientation is allowed. While for billiards without a reflection symmetry, for instance, the Africa billiard, one can consider a billiard of its mirror image, and for scars on one given orbit, the corresponding scar under parity operation has the reverse orientation. Note that our results can be generalized to more divergent physical pictures, e.g., particle-hole symmetry, negative potential, mirror reflection and their combinations, where the chirality still exists, although the spin behavior can be different. For the details of the system's behavior under symmetry operations, please refer to \ref{Appendix-D}.

Our complete understanding of the $\mathcal{T}$-breaking of the system leads to a control mechanism of the chiral scars, which can interchange chiral scars and non-chiral scars, although the applied magnetic flux for different scarring orbits can be different. This subtle $\mathcal{T}$-breaking phenomena by the odd periodic orbits and the edge states can have significant implications on the transport behavior and spin textures of the relativistic pseudoparticles \cite{XNLZ:2012}, or distinct magnetic response that could be applicable in quantum information devices, e.g., relativistic qubits \cite{FL:2013}. Our finding thus provides concrete grounds for both novel applications of the newly discovered 2D relativistic materials and the basic knowledge of relativistic quantum chaos.

\ack
We thank Dr. H.-Y. Xu, Prof. Y.-C. Lai and Prof. B. Dietz for helpful discussions. This work was supported by Grant No. 2015CB921503 from the MOST of China, by NNSF of China under Grants No. 11135001, No. 11375074, No. 11422541, No. 11504366, and No. 11434010, and by Doctoral Fund of Ministry of Education of China under Grant No. 20130211110008.

\appendix

\section{Circular Dirac A-B billiard with vanishing inner radius.}\label{Appendix-A}
To solve the chaotic Dirac A-B billiard with vanishing inner radius, we need to solve the eigenstates of the circular A-B billiard used as basis for conformal mapping. In particular, the system we shall study contains a single massless spin-half particle with charge $q$ confined by hard walls (infinite mass confinement) in a circular ring domain with inner radius $\xi \to 0$. The billiard system is threaded by a single line of magnetic flux $\Phi$ at the origin. We choose a non-divergent gauge in which the lines of the vector potential $\bi{A}$ are the contours of a scalar function $F(\bi{r})=-\ln(|\bi{r}|)$,
\begin{equation}
\bi{A}(\bi{r}) = \frac{\Phi}{2\pi} (\frac{\partial F}{\partial y}, -\frac{\partial F}{\partial x}) = \frac{\Phi}{2\pi} (-\frac{\sin\theta}{r}, \frac{\cos\theta}{r}),
\end{equation}
Note that $\nabla \cdot \bi{A} = 0$ and $\nabla \times \bi{A} = \hat{\bi{n}} \Phi \delta(\bi{r})$, $\hat{\bi{n}}$ is the unit vector normal to the $z$ plane.

The Dirac equation can be written as
\begin{equation}\label{eq:eigenAB}
v_F\hat{\bi{\sigma}}\cdot (\hat{\bi{p}} - \frac{q}{c}\bi{A}) \psi = E\psi,
\end{equation}
where $\psi=[\psi_1,\psi_2]^T$. And the boundary condition is
\begin{equation}\label{eq:boundarycondition}
\frac{\psi_2}{\psi_1}|_{\partial_D} =  ie^{i\widetilde{\theta}(s)},
\end{equation}
where $s$ is the arc length of the boundary, starting from the cross point of the boundary with positive $x$-axis; $\widetilde{\theta}(s)$ is the angle to the positive $x$-axis for the normal vector at $s$.
For a circularly symmetric ring boundary, we have
\begin{equation*}
[\hat{J}_z, \hat{H}] = 0,
\end{equation*}
where $\hat{J}_z = -i\hbar \partial _{\theta} + (\hbar /2){\hat{\sigma}_z}$ is the total angular momentum operator. We can choose the simultaneous eigenstates of $\hat{H}$ and $\hat{J_z}$:
\begin{equation*}
\hat{J}_z \psi = (l + 1/2)\hbar\psi.
\end{equation*}
So, the solutions of (\ref{eq:eigenAB}) has a general form that can be written as
\begin{equation}\label{eq:psi}
\psi(\bi{r}) =N
\left( \begin{array}{c}
\phi(r)\\
i\chi(r)e^{i\theta}
\end{array} \right)
e^{il\theta},
\end{equation}
where $l = 0, \pm1, \pm2,\cdots$ and $N$ is the normalization factor.

The Dirac equation in polar coordinate is
\begin{eqnarray}\label{eq:polarcoord}
&\left( \begin{array}{cc}
0 & e^{-i\theta}\Big(\frac{\partial}{\partial r} - \frac{i}{r}\frac{\partial}{\partial \theta} - \frac{\alpha}{r}\Big)\\
e^{i\theta}\Big(\frac{\partial}{\partial r} + \frac{i}{r}\frac{\partial}{\partial \theta} + \frac{\alpha}{r}\Big) & 0
\end{array} \right)
\psi(\bi{r})= i\mu\psi(\bi{r}),
\end{eqnarray}
where $\mu \equiv E/({\hbar v_F})$ and $\alpha \equiv (q\Phi)/(h c)$. Substituting Equation (\ref{eq:psi}) into Equation (\ref{eq:polarcoord}), we can get
\begin{equation}\label{eq:polarsimp}
\left( \begin{array}{cc}
- \mu & \frac{d}{dr} + \frac{l + 1 - \alpha}{r}\\
-\frac{d}{dr} + \frac{l - \alpha}{r} & -\mu
\end{array} \right)
\left( \begin{array}{c}
\phi(r)\\
\chi(r)
\end{array} \right)
= 0.
\end{equation}

By canceling $\chi$ in Equation (\ref{eq:polarsimp}), we get the Bessel's differential equation
\begin{equation}
\Big( \frac{d^2}{dR^2} + \frac{1}{R}\frac{d}{dR} + 1 - \frac{(l - \alpha)^2}{R^2}   \Big) \phi(r) = 0,
\end{equation}
where $R = \mu r$ and $\nu = l - \alpha$. $\phi(r)$ can be wrritten as a linear combination of the Bessel function of the first kind $J_{\nu}(R)$ and the Bessel function of the second kind $N_{\nu}(R)$, i.e.,
\begin{equation}\label{eq:phi}
\phi(R) = J_{\nu}(R) + \beta N_{\nu}(R),
\end{equation}
where $\beta$ is a coefficient and can be determined by the boundary conditions. $\chi(R)$ satisfies the following equation
\begin{equation*}
-R\chi(R) = R\phi^{'}(R) - \nu \phi(R).
\end{equation*}
Employing the recursive relation of Bessel functions, we obtain
\begin{equation}\label{eq:kai}
\chi(R) = J_{\nu + 1}(R) + \beta N_{\nu + 1}(R).
\end{equation}

Then the inner and outer boundary conditions lead to
\begin{equation}
\left \{ \begin{array}{ll}
J_{\nu}(\mu \xi) + \beta N_{\nu}(\mu \xi) = -\big(J_{\nu + 1}(\mu \xi) + \beta N_{\nu +1 }(\mu \xi)\big), \\
J_{\nu}(\mu) + \beta N_{\nu}(\mu) = J_{\nu + 1}(\mu) + \beta N_{\nu +1 }(\mu).
\end{array} \right.
\end{equation}
By solving the above equations, $\beta$ is given as
\begin{equation}\label{eq:beta}
\beta = -\frac{J_{\nu + 1}(\mu \xi) + J_{\nu}(\mu \xi)}{N_{\nu + 1}(\mu \xi) + N_{\nu}(\mu \xi)} = -\frac{J_{\nu + 1}(\mu) - J_{\nu}(\mu)}{N_{\nu + 1}(\mu) - N_{\nu}(\mu)},
\end{equation}
where the eigenvalue $\mu$ (and thus $E={\hbar v_F}\mu$) can be obtained by solving the equation
\begin{eqnarray}\label{eq:mu}
\big[J_{\nu + 1}(\mu) - J_{\nu}(\mu)\big]\cdot\big[N_{\nu + 1}(\mu \xi) + N_{\nu}(\mu \xi)\big] \nonumber\\
=\big[J_{\nu + 1}(\mu \xi) + J_{\nu}(\mu \xi)\big]\cdot\big[N_{\nu + 1}(\mu) - N_{\nu}(\mu)\big].
\end{eqnarray}
Equation (\ref{eq:mu}) can be simplified by the special properties of the Bessel functions listed below, i.e.,
\begin{eqnarray*}
\lim_{x \rightarrow 0} J_{\lambda}(x) \sim \frac{x^{\lambda}}{2^{\lambda}\Gamma(1+ \lambda)}=
\cases{
0  &\text{($\lambda > 0$ and $\lambda = -integer$),}\cr
1  &\text{($\lambda = 0$)},\cr
\infty  &\text{($\lambda<0$ and $\lambda\ne - integer$).}} \\
N_{\lambda}(x)=\frac{\cos(\lambda \pi) J_{\lambda}(x) - J_{-\lambda}(x)} {\sin(\lambda \pi)},
\quad \lim_{x \rightarrow 0} N_{\lambda}(x) \sim \infty.
\end{eqnarray*}

For $\nu$ being an integer, the right hand side of Equation (\ref{eq:mu}) is finite. Since both $N_{\nu + 1}(\mu \xi)$ and $N_{\nu}(\mu \xi)$ diverge as $\xi$ goes to zero, $(J_{\nu + 1}(\mu) - J_{\nu}(\mu))$ must be zero. So we have

(1). $\nu=$ integer,
\begin{eqnarray}\label{eq:mu1}
J_{\nu}(\mu) \approx J_{\nu+1}(\mu).
\end{eqnarray}

For $\nu$ not being an integer, $N_{\nu}$ can be expressed as a linear combination of $J_{\nu}$ and $J_{-\nu}$, so Equation (\ref{eq:mu}) can be simplified as
\begin{eqnarray*}
\big[J_{\nu + 1}(\mu) - J_{\nu}(\mu)\big]\cdot\big[J_{-(\nu + 1)}(\mu \xi) - J_{-\nu}(\mu \xi)\big] \nonumber\\
=\big[J_{\nu + 1}(\mu \xi) + J_{\nu}(\mu \xi)\big]\cdot\big[J_{-(\nu + 1)}(\mu) + J_{-\nu}(\mu)\big].
\end{eqnarray*}
In the $\xi \to 0$ limit, we can get

(2). $\nu>0$, $J_{-(\nu +1)}(\mu\xi) - J_{-\nu}(\mu\xi)\to \infty$, $J_{\nu +1}(\mu\xi) + J_{\nu}(\mu\xi) \to 0$, thus
\begin{eqnarray}
J_{\nu}(\mu) \approx J_{\nu + 1}(\mu).
\end{eqnarray}

(3). $\nu < -1$, $J_{-(\nu + 1)}(\mu \xi) - J_{-\nu}(\mu\xi) \to 0$, $J_{\nu + 1}(\mu\xi) + J_{\nu}(\mu\xi) \to \infty$, yielding
\begin{eqnarray}
J_{-\nu}(\mu) \approx -J_{-(\nu + 1)}(\mu).
\end{eqnarray}

For $-1 < \nu <0$, $J_{-(\nu + 1)}(\mu\xi) \sim \frac{\xi^{-(1+\nu)}}{\Gamma(-\nu)}$, $J_{\nu}(\mu\xi) \sim \frac{\xi^{\nu}}{\Gamma(1 + \nu)}$, $J_{-\nu}(\mu\xi)\to 0$, $J_{\nu + 1}(\mu\xi)\to 0$. So, we have

(4). $-1/2<\nu<0$, $J_{\nu + 1}(\mu\xi)/J_{\nu}(\mu\xi)\sim \xi^{-1-2\nu}\Gamma(1+\nu)/\Gamma(-\nu)\to +\infty$, thus
\begin{eqnarray}
J_{\nu}(\mu) \approx J_{\nu + 1}(\mu).
\end{eqnarray}

(5). $-1<\nu<-1/2$, $J_{-(\nu + 1)}(\mu\xi)/J_{\nu}(\mu\xi)\sim \xi^{-1-2\nu}\Gamma(1+\nu)/\Gamma(-\nu)\to 0$, so
\begin{eqnarray}
J_{-\nu}(\mu) \approx -J_{-(\nu + 1)}(\mu).
\end{eqnarray}

(6). $\nu = -1/2$, $N_{-\frac{1}{2}}(x) = J_{\frac{1}{2}}(x)$, $N_{\frac{1}{2}}(x) = -J_{-\frac{1}{2}}(x)$.
Using Equation~(\ref{eq:beta}), we can get $\beta = 1$, $J_{\nu + 1}(\mu) - J_{\nu}(\mu) = J_{\nu + 1}(\mu) + J_{\nu}(\mu)$, thus
\begin{eqnarray}\label{eq:mu6}
J_{-\frac{1}{2}}(\mu) \approx 0.
\end{eqnarray}

For the simplified equations (\ref{eq:mu1})-(\ref{eq:mu6}), we can get the eigenvalues $\mu_{lm}(\alpha)$, where the magnetic flux $\alpha$ can be regarded as a control parameter, $l$ and $\nu$ are related by $\nu=l-\alpha$, and $m$ represents the $m$th solution for a given $l$.

Once the $\mu_{lm}(\alpha)$ is obtained, substituting it back into Equation (\ref{eq:beta}), we can get the corresponding $\beta_{lm}(\alpha)$. Substituting these two quantities back to equations (\ref{eq:psi}), (\ref{eq:phi}) and (\ref{eq:kai}), we can obtain the corresponding eigenfunction $\psi_{lm}(\alpha)$:
\begin{eqnarray}\label{eq:psic}
&\psi_{lm}(\bi{r},\alpha)=N_{lm}
\left( \begin{array}{c}
\phi_{lm}(r)\\
i\chi_{lm}(r)e^{i\theta}
\end{array} \right)
e^{il\theta} \\
&=N_{lm}
\left( \begin{array}{c}
J_{\nu}(\mu_{lm} r)+\beta_{lm} N_{\nu}(\mu_{lm} r) \\
i\big(J_{\nu+1}(\mu_{lm} r)+\beta_{lm} N_{\nu+1}(\mu_{lm} r)\big)e^{i\theta}
\end{array} \right)
e^{il\theta}. \nonumber
\end{eqnarray}

In particular, we can get the simplified expressions for the eigenfunctions by appropriate approximations as following.

(1). $\nu$ is an integer:

Note that the divergence property of $N_{\nu}(x)$ is as follows,
\begin{eqnarray*}
\lim_{x \rightarrow 0} N_{0}(x) & \sim \frac{2}{\pi}\ln \frac{x}{2} \vert_{x \to 0}\\
\lim_{x \rightarrow 0} N_{\nu}(x) &\sim \frac{-(\nu-1)!}{\pi}\Big(\frac{x}{2}\Big)^{-\nu} \vert_{x \to 0} \quad \nu = 1,2,\cdots\\
N_{-\nu}(x) &= (-1)^{\nu}N_{\nu}(x)
\end{eqnarray*}
So, if $\nu$ is non-negative and $r \ge \xi$,
\begin{eqnarray*}
\beta &\approx -\frac{J_{\nu}(\mu\xi)}{N_{\nu+1}(\mu\xi)}.
\end{eqnarray*}
Then
\begin{eqnarray*}
\beta N_{\nu}(\mu r) &= -\frac{J_{\nu}(\mu\xi)}{N_{\nu+1}(\mu\xi)}N_{\nu}(\mu r) \approx 0,
\end{eqnarray*}
as $N_{\nu+1}$ diverges faster than $N_{\nu}$ at $r = \xi$. We also have
\begin{eqnarray*}
\beta N_{\nu+1}(\mu r) &= -\frac{J_{\nu}(\mu\xi)}{N_{\nu+1}(\mu\xi)}N_{\nu+1}(\mu r).
\end{eqnarray*}
Thus
\begin{eqnarray}\label{eq:wavefunction1}
\left \{ \begin{array}{ll}
\phi_{lm}(r) &= J_{\nu}(\mu_{lm} r),\\
\chi_{lm}(r) &= J_{\nu+1}(\mu_{lm} r) -\frac{J_{\nu}(\mu_{lm}\xi)}{N_{\nu+1}(\mu_{lm}\xi)}N_{\nu+1}(\mu_{lm} r).
\end{array} \right.
\end{eqnarray}
Note that the term $-\frac{J_{\nu}(\mu_{lm}\xi)}{N_{\nu+1}(\mu_{lm}\xi)}N_{\nu+1}(\mu_{lm} r)$
has little influence on the eigenvalues and the eigenfunctions. The reason is that for $r \gg \xi$, $N_{\nu+1}(\mu_{lm} r)$ is finite, while $J_{\nu}(\mu_{lm}\xi)/N_{\nu+1}(\mu_{lm}\xi) \sim -[\pi/(2^{\nu+1/2}\nu !)^2](\mu_{lm}\xi)^{2\nu+1} \sim -\xi^{2\nu+1} \sim 0$, thus the whole term approaches zero. When $r=\xi$, this term becomes $-J_{\nu}(\mu_{lm}\xi) \sim -(\mu_{lm}\xi)^{\nu}/ (2^{\nu}\nu!)$, and is finite. However, this term guarantees the boundary condition $\chi_{lm}(\mu_{lm}\xi)/\phi_{lm}(\mu_{lm}\xi)=-1$ (Equation \ref{eq:boundarycondition}) at the inner boundary $r=\xi$ and leads to a clockwise flow at this boundary.

If $\nu$ is negative,
\begin{eqnarray*}
\beta \approx -\frac{J_{\nu+1}(\mu\xi)}{N_{\nu}(\mu\xi)},\quad
\beta N_{\nu}(\mu r) = -\frac{J_{\nu+1}(\mu\xi)}{N_{\nu}(\mu\xi)}N_{\nu}(\mu r),\\
\beta N_{\nu+1}(\mu r) = -\frac{J_{\nu+1}(\mu\xi)}{N_{\nu}(\mu\xi)}N_{\nu+1}(\mu r) \approx 0 ,
\end{eqnarray*}
thus
\begin{eqnarray}
\left \{ \begin{array}{ll}
\phi_{lm}(r) &= J_{\nu}(\mu_{lm} r)-\frac{J_{\nu+1}(\mu_{lm}\xi)}{N_{\nu}(\mu_{lm}\xi)}N_{\nu}(\mu_{lm} r),\\
\chi_{lm}(r) &= J_{\nu+1}(\mu_{lm} r).
\end{array} \right.
\end{eqnarray}
Similarly, the term $-\frac{J_{\nu+1}(\mu_{lm}\xi)}{N_{\nu}(\mu_{lm}\xi)}N_{\nu}(\mu_{lm} r)$
has little effect on the eigenvalues and eigenfunctions but guarantees the boundary condition at the inner circle. Note that if $\nu$ is an integer, $J_{\nu}(\mu_{lm}r) = (-1)^{\nu}J_{-\nu}(\mu_{lm}r)$, and further simplification can be obtained.

For $\nu$ not being an integer, the asymptotic behavior of the first class Bessel Function is
\begin{eqnarray*}
\lim_{x \rightarrow 0} J_{\nu}(x) \sim \frac{x^{\nu}}{2^{\nu} \Gamma(1+\nu)}\vert_{x \rightarrow 0},
\end{eqnarray*}
based on which we can get the following approximations.

(2). $\nu > 0$ and $\nu$ is not an integer:
\begin{eqnarray*}
N_{\nu}(\mu\xi) \approx -\frac{1}{\sin\nu\pi}\cdot J_{-\nu}(\mu\xi),\quad
N_{\nu + 1}(\mu\xi)\approx \frac{1}{\sin\nu\pi}\cdot J_{-(\nu + 1)}(\mu\xi),\\
\beta= -\frac{\sin\nu\pi \cdot [J_{\nu+1}(\mu\xi) + J_{\nu}(\mu\xi)]}{J_{-(\nu+1)}(\mu\xi) - J_{-\nu}(\mu\xi)}
\approx -\frac{\sin\nu\pi \cdot J_{\nu}(\mu\xi)}{J_{-(\nu + 1)}(\mu\xi)},\\
\beta \cdot N_{\nu}(\mu r)\approx \frac{J_{\nu}(\mu\xi)}{J_{-(\nu+1)}(\mu\xi)}\cdot  J_{-\nu}(\mu r) \approx 0,\\
\beta \cdot N_{\nu + 1}(\mu r) \approx -\frac{J_{\nu}(\mu\xi)}{J_{-(\nu+1)}(\mu\xi)}\cdot J_{-(\nu+1)}(\mu r).
\end{eqnarray*}
Thus
\begin{eqnarray}
\left \{ \begin{array}{ll}
\phi_{lm}(r) &\approx J_{\nu}(\mu_{lm} r),\\
\chi_{lm}(r) &\approx J_{\nu+1}(\mu_{lm} r) - \frac{J_{\nu}(\mu_{lm}\xi)}{J_{-(\nu+1)}(\mu_{lm}\xi)}\cdot J_{-(\nu+1)}(\mu_{lm} r).
\end{array} \right.
\end{eqnarray}
Note that for $r > \xi$, the second term in $\chi_{lm}(r)$ approaches to zero, so $\chi_{lm}(r) \approx J_{\nu+1}(\mu_{lm} r)$. While for $r \rightarrow \xi \rightarrow 0$, $\chi_{lm}(r) \approx -J_{\nu}(\mu_{lm}\xi)$, which satisfies the boundary condition Equation (\ref{eq:boundarycondition}) at $r = \xi$ and leads to a clockwise current at the inner boundary.

(3). $\nu<-1$ and $\nu$ is not an integer.

If $\nu$ is not a half-integer,
\begin{eqnarray*}
\beta \approx -\tan\nu\pi\big(1 - \frac{1}{\cos\nu\pi}\cdot \frac{J_{-(\nu+1)}(\mu\xi)}{J_{\nu}(\mu\xi)}\big),\\
\beta \cdot N_{\nu}(\mu r) \approx -J_{\nu}(\mu r) + \frac{1}{\cos\nu\pi}\cdot J_{-\nu}(\mu r)  + \frac{1}{\cos\nu\pi} \frac{J_{-(\nu+1)}(\mu\xi)}{J_{\nu}(\mu\xi)}\cdot J_{\nu}(\mu r),\\
\beta \cdot N_{\nu+1}(\mu r) = -J_{\nu+1}(\mu r) - \frac{1}{\cos\nu\pi}\cdot J_{-(\nu+1)}(\mu r).
\end{eqnarray*}
Thus
\begin{eqnarray}\label{eq:nu-minus1}
\left \{ \begin{array}{ll}
\phi_{lm}(r) & \approx J_{-\nu}(\mu_{lm} r) + \frac{J_{-(\nu+1)}(\mu_{lm}\xi)}{J_{\nu}(\mu_{lm}\xi)}J_{\nu}(\mu_{lm} r),\\
\chi_{lm}(r) &\approx -J_{-(\nu+1)}(\mu_{lm} r).
\end{array} \right.
\end{eqnarray}

If $\nu$ is a half-integer,
\begin{eqnarray*}
N_{\nu}(\mu\xi) \approx -\frac{1}{\sin\nu\pi}\cdot J_{-\nu}(\mu\xi),\quad
N_{\nu + 1}(\mu\xi)\approx \frac{1}{\sin\nu\pi}\cdot J_{-(\nu + 1)}(\mu\xi),\\
\beta= -\frac{\sin\nu\pi \cdot [J_{\nu+1}(\mu\xi) + J_{\nu}(\mu\xi)]}{J_{-(\nu+1)}(\mu\xi) - J_{-\nu}(\mu\xi)} \approx -\frac{\sin\nu\pi \cdot J_{\nu}(\mu\xi)}{J_{-(\nu + 1)}(\mu\xi)},\\
\beta \cdot N_{\nu}(\mu r)\approx \frac{J_{\nu}(\mu\xi)}{J_{-(\nu+1)}(\mu\xi)}\cdot J_{-\nu}(\mu r),\\
\beta \cdot N_{\nu + 1}(\mu r) \approx -\frac{J_{\nu}(\mu\xi)}{J_{-(\nu+1)}(\mu\xi)}\cdot J_{-(\nu+1)}(\mu r).
\end{eqnarray*}
Thus if we take the term ${J_{\nu}(\mu\xi)}/{J_{-(\nu+1)}(\mu\xi)}$ into the normalization factor, we can get the same formula as Equation~({\ref{eq:nu-minus1}})
\begin{eqnarray}
\left \{ \begin{array}{ll}
\phi_{lm}(r) & \approx J_{-\nu}(\mu_{lm} r) + \frac{J_{-(\nu+1)}(\mu_{lm}\xi)}{J_{\nu}(\mu_{lm}\xi)}J_{\nu}(\mu_{lm} r),\\
\chi_{lm}(r) &\approx -J_{-(\nu+1)}(\mu_{lm} r).
\end{array} \right.
\end{eqnarray}

(4). $-1/2<\nu<0$:
\begin{eqnarray*}
N_{\nu}(\mu\xi) &= \frac{\cos\nu\pi}{\sin\nu\pi}\cdot J_{\nu}(\mu\xi) - \frac{1}{\sin\nu\pi}\cdot J_{-\nu}(\mu\xi)\approx \frac{\cos\nu\pi}{\sin\nu\pi}\cdot J_{\nu}(\mu\xi), \\
N_{\nu+1}(\mu\xi) &= \frac{\cos(\nu+1)\pi}{\sin(\nu+1)\pi}\cdot J_{\nu+1}(\mu\xi)
  - \frac{1}{\sin(\nu+1)\pi}\cdot J_{-(\nu+1)}(\nu\xi) \nonumber\\
&\approx \frac{1}{\sin\nu\pi}\cdot J_{-(\nu+1)}(\mu\xi),
\end{eqnarray*}
\begin{eqnarray*}
\beta = -\frac{J_{\nu+1}(\mu\xi)+J_{\nu}(\mu\xi)}{\frac{1}{\sin\nu\pi}\big[\cos\nu\pi \cdot J_{\nu}(\mu\xi) + J_{-(\nu+1)}(\mu\xi)\big]}
\approx -\frac{\sin\nu\pi\cdot J_{\nu}(\mu\xi)}{\cos\nu\pi \cdot J_{\nu}(\mu\xi) + J_{-(\nu+1)}(\mu\xi)},
\end{eqnarray*}
\begin{eqnarray*}
\beta\cdot N_{\nu}(\mu r) &= -\frac{J_{\nu}(\mu\xi)}{J_{-(\nu + 1)}(\mu\xi)}\cdot \big[\cos\nu\pi \cdot J_{\nu}(\mu r) - J_{-\nu}(\mu r)\big]\approx 0, \\
\beta\cdot N_{\nu+1}(\mu r) &\approx -\frac{J_{\nu}(\mu\xi)}{J_{-(\nu + 1)}(\mu\xi)}\cdot J_{-(\nu + 1)}(\mu r).
\end{eqnarray*}
Thus
\begin{eqnarray}
\left \{ \begin{array}{ll}
\phi_{lm}(r) &\approx J_{\nu}(\mu_{lm} r) + \frac{J_{\nu}(\mu_{lm}\xi)}{J_{-(\nu + 1)}(\mu_{lm}\xi)}\cdot J_{-\nu}(\mu_{lm} r), \\
\chi_{lm}(r) &\approx J_{\nu+1}(\mu_{lm} r) - \frac{J_{\nu}(\mu\xi)}{J_{-(\nu+1)}(\mu_{lm}\xi)}\cdot J_{-(\nu+1)}(\mu_{lm} r).
\end{array} \right.
\end{eqnarray}

(5). $-1<\nu<-1/2$:
\begin{eqnarray*}
N_{\nu}(\mu\xi)\approx \frac{\cos\nu\pi}{\sin\nu\pi}\cdot J_{\nu}(\mu\xi), \quad
N_{\nu+1}(\mu\xi) \approx \frac{1}{\sin\nu\pi}\cdot J_{-(\nu+1)}(\mu\xi), \\
\beta \approx -\tan\nu\pi(1 - \frac{1}{\cos\nu\pi}\cdot \frac{J_{-(\nu+1)}(\mu\xi)}{J_{\nu}(\mu\xi)}),\\
\beta \cdot N_{\nu}(\mu r) \approx  -J_{\nu}(\mu r) + \frac{1}{\cos\nu\pi}\cdot \big[J_{-\nu}(\mu r) + \frac{J_{-(\nu+1)}(\mu r)}{J_{\nu}(\mu\xi)}\cdot J_{\nu}(\mu r)\big],\\
\beta \cdot N_{\nu+1}(\mu r)= -J_{\nu + 1}(\mu\xi) - \frac{1}{\cos\nu\pi}\cdot \big(J_{-(\nu+1)}(\mu r) + \frac{J_{-(\nu + 1)}(\mu\xi)}{J_{\nu}(\mu\xi)}\cdot J_{\nu+1}(\mu r)\big).
\end{eqnarray*}
Thus
\begin{eqnarray}
\left \{ \begin{array}{ll}
\phi_{lm}(r)=& \frac{J_{-(\nu+1)}(\mu_{lm}\xi)}{J_{\nu}(\mu_{lm}\xi)}\cdot J_{\nu}(\mu_{lm} r) + J_{-\nu}(\mu_{lm} r),\\
\chi_{lm}(r)=& -J_{-(\nu + 1)}(\mu_{lm} r)+ \frac{J_{-(\nu+1)}(\mu_{lm}\xi)}{J_{\nu}(\mu_{lm}\xi)}\cdot J_{\nu+1}(\mu_{lm} r).
\end{array} \right.
\end{eqnarray}

(6). $\nu = -1/2$, then we have $\beta \approx 1$,
thus
\begin{eqnarray}\label{eq:wavefunction6}
\left \{ \begin{array}{ll}
\phi_{lm}(r) &\approx J_{-1/2}(\mu_{lm} r) + J_{1/2}(\mu_{lm} r),\\
\chi_{lm}(r) &\approx J_{1/2}(\mu_{lm} r) - J_{-1/2}(\mu_{lm} r).
\end{array} \right.
\end{eqnarray}

Substituting equations (\ref{eq:wavefunction1})-(\ref{eq:wavefunction6}) back to Equation (\ref{eq:psic}), we get the simplified eigenfunctions $\psi_{lm}(\mathbf{r},\alpha)$ corresponding to eigenvalues $\mu_{lm}(\alpha)$. Note that when we calculate the eigenfunctions numerically, we can ignore the small value terms and use the following approximations:\\
(1). $\nu>-1/2$,
\begin{eqnarray}
\left \{ \begin{array}{ll}
\phi_{lm}(r) &= J_{\nu}(\mu_{lm} r),\\
\chi_{lm}(r) &= J_{\nu+1}(\mu_{lm} r).
\end{array} \right.
\end{eqnarray}\\
(2). $\nu<-1/2$,
\begin{eqnarray}
\left \{ \begin{array}{ll}
\phi_{lm}(r) & \approx J_{-\nu}(\mu_{lm} r),\\
\chi_{lm}(r) &\approx -J_{-(\nu+1)}(\mu_{lm} r).
\end{array} \right.
\end{eqnarray}\\
(3). $\nu=-1/2$,
\begin{eqnarray}
\left \{ \begin{array}{ll}
\phi_{lm}(r) &\approx J_{-1/2}(\mu_{lm} r) + J_{1/2}(\mu_{lm} r),\\
\chi_{lm}(r) &\approx J_{1/2}(\mu_{lm} r) - J_{-1/2}(\mu_{lm} r).
\end{array} \right.
\end{eqnarray}

\section{Physical process of each local reflection} \label{Appendix-B}

In this section, by employing the model of plane wave reflection at a straight potential, we shall show that the wave at the boundary is an eigenfunction for $\hat{S}_y$ with an eigenvalue of $\hbar/2$, regardless of the incident angle. That is, whether the incident wave is coming upwards or coming downwards, the spin always points up (or counterclockwise), indicating chirality.
Therefore, a time-reversed wave will not result in a time-reversed spin polarization at the boundary, leading to $\mathcal{T}$-breaking.

The origin of spin polarization can be understood by analyzing the phase change at each reflection. We found that for each local reflection, the difference for the phase change during a reflection and its time-reversed counterpart has an additional $\pi$ contribution, which is also an indication of $\mathcal{T}$-breaking. Therefore, each reflection at the boundary breaks the time-reversal symmetry.

With these results, we further provide a complementary understanding of the global phase change difference of even and odd closed orbits discussed by Berry {\it et al.} \cite{Berry:1987}.

\begin{figure}[h]
\centering
  \includegraphics[width=0.9\textwidth]{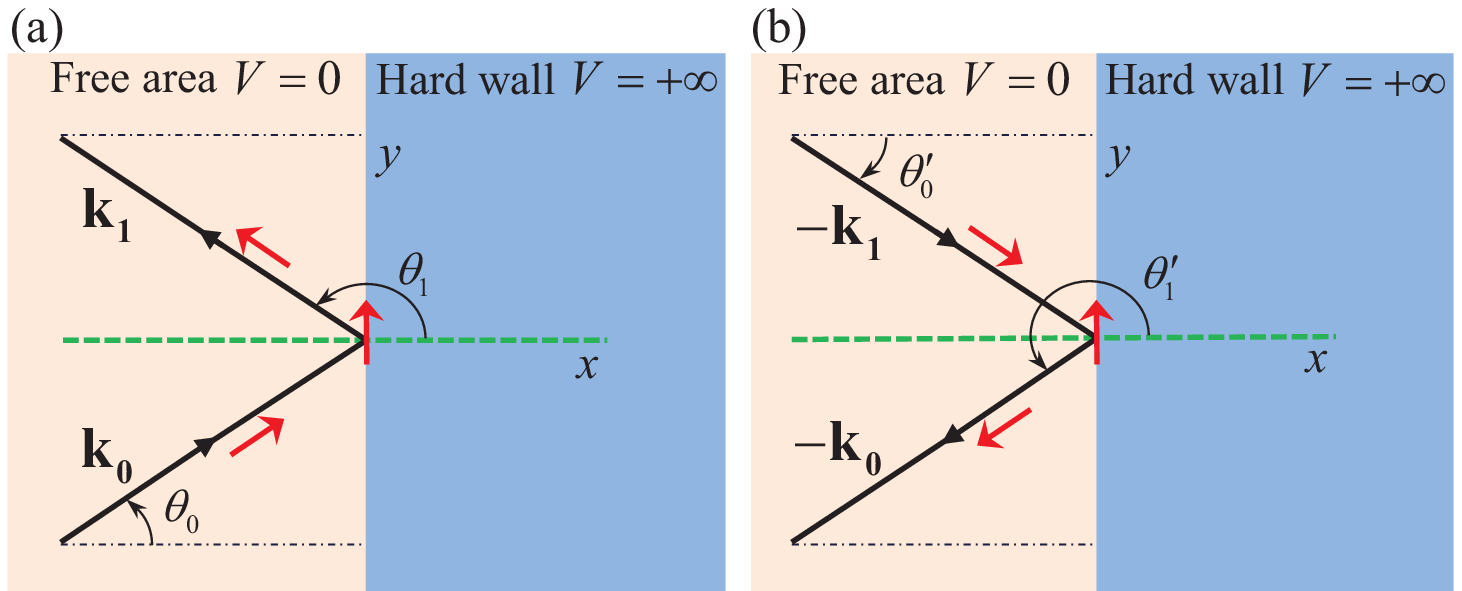}
\caption{Incident and reflected plane waves (black arrow) and the spin (red arrow) corresponding to the superposition of waves at the boundary.}
  \label{fig:planewave}
\end{figure}

To gain insight into the boundary effect on the spin, we employ the model of plane-wave reflection on a straight boundary, which has been discussed in details in \cite{Berry:1987,XHLG:2015}, the schematic diagram is shown in figure~\ref{fig:planewave} (for generality we take $V>E$ in area 2). Here we briefly list their results as Equations (\ref{eq:planewave}-\ref{eq:Transmission}). The wave (incident plus reflected) in the plain area can be written as 
\begin{eqnarray} \label{eq:planewave}
\Psi_1 = & \frac{1}{\sqrt{2}}\Bigg[
\left( \begin{array}{c}
\exp{\{-\frac{1}{2}i\theta_0\}}\\
\exp{\{\frac{1}{2}i\theta_0\}}
\end{array} \right)\exp{\{i\bi{k}_0\cdot \bi{r}\}}
\nonumber \\
& + R
\left( \begin{array}{c}
\exp{\{-\frac{1}{2}i\theta_1\}}\\
\exp{\{\frac{1}{2}i\theta_1\}}
\end{array} \right)
\exp{\{i\bi{k}_1\cdot \bi{r}\}}\Bigg],
\end{eqnarray}
and the transmitted wave in the potential area is
\begin{eqnarray}\label{eq:transwave}
\Psi_2 = \frac{T}{\sqrt{2}}
\left( \begin{array}{c}
-i\lambda_1\\
\lambda_2
\end{array} \right)
e^{-qx}e^{iKy},
\end{eqnarray}
where $R$, $T$ are the reflection and transmission coefficients, respectively, the incident wave vector $\bi{k}_0 = (k\cos\theta_0, k\sin\theta_0)$ and the reflected wave vector $\bi{k}_1 = (k\cos\theta_1, k\sin\theta_1)$, $K = k\sin\theta_0$ and $q = \sqrt{\frac{V^2 - E^2}{\hbar^2v_F^2} + K^2}$; $E = \hbar v_F k$, and $\lambda_1 = \sqrt{\frac{(V + E)(q - K)}{Vq - EK}}$; $\lambda_2 = \sqrt{\frac{(V - E)(q + K)}{Vq - EK}}$.
Matching the two waves at $x = 0$ and using the convention to relate the incident and reflected directions by specularity \cite{Berry:1987}
\begin{eqnarray} \label{eq:specularity}
\theta_1 = \pi + 2\widetilde{\theta}(s) - \theta_0,
\end{eqnarray}
where $\widetilde{\theta}(s)=0$ for the special case we consider here. We can obtain
\begin{eqnarray}\label{eq:R-coe}
R = \frac{i\lambda - e^{i\theta_0}}{i - \lambda e^{i\theta_0}} = e^{i(2\gamma + \theta_0 - \frac{\pi}{2})},
\end{eqnarray}
where the parameters $\gamma$ and $\lambda$ are defined through
\begin{eqnarray}
\tan\gamma &= \frac{\lambda_1 - \lambda_2\sin\theta_0}{\lambda_2\cos\theta_0} = \frac{1 - \lambda\sin\theta_0}{\lambda\cos\theta_0}, \\
\lambda = \frac{\lambda_2}{\lambda_1} &= \sqrt{\frac{(V - E)(q + K)}{(V + E)(q - K)}} = \frac{V - E}{\hbar v_F (q - K)} = \frac{\hbar v_F(q + K)}{V + E}.
\end{eqnarray}
Also, the transmission coefficient is given by
\begin{eqnarray}\label{eq:Transmission}
T = \frac{2\cos \gamma}{\lambda_2}e^{i(\gamma + \frac{\theta_0}{2})}.
\end{eqnarray}
Note that the above convention in Equation~(\ref{eq:specularity}) actually implies the change from $\theta_0$ to $\theta_1$ by rotating the angle counterclockwisely (figure \ref{fig:planewave}). If the change of the angle is made by rotating clockwisely, there will be an additional $2\pi$ at the right side of Equation~(\ref{eq:specularity}), and an additional phase $\pi$ in the plane wave in the second term of Equation~(\ref{eq:planewave}) for there is a prefactor $1/2$. But the final results are unchanged. Another important property of the refection coefficient $R$ is that $R(\theta_0) = R(-\theta_0)$, i.e., it is the same for the forward or backward incidence. For finite $V > E$, $R$ is not a constant but a position dependent function. And as $V$ goes to infinity, $R$ becomes 1. In the following unless otherwise specified we assume $E>0$, $V \to \infty$ and $R=1$.

{\bf Spin orientation.} The wave-function on the boundary in figure \ref{fig:planewave}(a) is
\begin{eqnarray*}
\Psi_1 =
\left( \begin{array}{c}
\exp{\{-\frac{1}{2}i\theta_0\}} + \exp{\{-\frac{1}{2}i\theta_1\}}\\
\exp{\{\frac{1}{2}i\theta_0\}} + \exp{\{\frac{1}{2}i\theta_1\}}
\end{array} \right) = \left( \begin{array}{c}
\exp{\{-\frac{1}{2}i\theta_0\}} - i\exp{\{\frac{1}{2}i\theta_0\}}\\
\exp{\{\frac{1}{2}i\theta_0\}} + i\exp{\{-\frac{1}{2}i\theta_0\}}
\end{array} \right),
\end{eqnarray*}
while in figure \ref{fig:planewave}(b),
\begin{eqnarray*}
\Psi_2 =
\left( \begin{array}{c}
\exp{\{-\frac{1}{2}i\theta_{0}^{'}\}} + \exp{\{-\frac{1}{2}i\theta_{1}^{'}\}}\\
\exp{\{\frac{1}{2}i\theta_{0}^{'}\}} + \exp{\{\frac{1}{2}i\theta_{1}^{'}\}}
\end{array} \right) = \left( \begin{array}{c}
\exp{\{-\frac{1}{2}i\theta_{0}^{'}\}} - i\exp{\{\frac{1}{2}i\theta_{0}^{'}\}}\\
\exp{\{\frac{1}{2}i\theta_{0}^{'}\}} + i\exp{\{-\frac{1}{2}i\theta_{0}^{'}\}}
\end{array} \right).
\end{eqnarray*}
The $y$-direction spin operator is $\hat{S}_y=(\hbar/2)\hat{\sigma}_y$. It is straightforward to verify that both $\Psi_1$ and $\Psi_2$ are
eigen-functions of $\hat{S}_y$, with the same eigen-value $\hbar/2$:
\begin{eqnarray} \label{eq:SyeigenEV}
\hat{S}_y\Psi_{1,2} = \frac{\hbar}{2}\Psi_{1,2}.
\end{eqnarray}
That is, the two opposite incident cases have the same spin orientation on the boundary!

We can see the spin always points to the counterclockwise direction regardless of the incident angle (this also can be seen by the boundary condition). This indicates that each local collision breaks the $\mathcal{T}$-symmetry due to the interaction between spin and the infinite mass boundary.

Note that although in the configuration space the probability
density current on the boundary have the same orientation for the two opposite incident directions, the magnitudes are typically different \cite{XHLG:2015}.

{\bf Additional $\pi$ phase for reversed reflection.} Now, we can carefully study the phase change of the spinor wavefunction during one reflection under the special condition $V\to +\infty$ and the corresponding $R = 1$. Suppose the incident angle is $\theta_0$ and the reflected angle is $\theta_1$, as shown in figure~\ref{fig:planewave}(a). These two angles are related by Equation~(\ref{eq:specularity}). So, according to Equation~(\ref{eq:planewave}) the phase difference between these two directions can be written as
\begin{eqnarray}\label{eq:localangle1}
\delta_{+} = \frac{1}{2}(\theta_1 - \theta_0) = \frac{1}{2}(\pi + 2\widetilde{\theta}(s) - 2\theta_0),
\end{eqnarray}
If we reverse the reflection direction, the incident and reflected angles are labeled as $\theta^{'}_{0}$ and $\theta^{'}_{1}$, where $\theta^{'}_{0} = -\theta_0 + 2\widetilde{\theta}(s) +2n\pi$, $n$ is an integer, as shown in figure~\ref{fig:planewave}(b). The phase difference between these two directions is
\begin{eqnarray}\label{eq:localangle2}
\delta_{-} &= \frac{1}{2}(\theta^{'}_{1} - \theta^{'}_{0}) = \frac{1}{2}(\pi + 2\widetilde{\theta}(s) - 2\theta^{'}_{0}) \nonumber = -\delta_{+} + \pi - 2n\pi,
\end{eqnarray}
Note that $\widetilde{\theta}(s) = 0$ and the additional $2n\pi$ has no observable effect here, which can be ignored. For each collision, the phase change of a pair of two opposite incident directions changes a minus sign as well as an additional phase $\pi$, which ensures the spin polarization at the boundary.

{\bf Global phase change.} Now let us consider the global phase changes based on the local phase change relation for each reflection. For a closed orbit with the initial incident angle $\theta_0$ based on Equation~(\ref{eq:localangle1}), closure means that the whole phase change is
\begin{eqnarray}\label{eq:positivephase}
\Delta_{+} = \frac{1}{2}(\theta_{n} - \theta_{0}) = m\pi,
\end{eqnarray}
where $m$ is an integer. If we reverse the initial direction of the same orbit based on Equation~(\ref{eq:localangle2}), the global phase change satisfies
\begin{eqnarray}\label{eq:negativephase}
\Delta_{-} = \frac{1}{2}(\theta^{'}_{n} - \theta^{'}_{0}) = -\Delta_{+} + N\pi,
\end{eqnarray}
where $N$ is the total number of reflections. So the phase difference between the two reversed orbits caused by boundary is
\begin{eqnarray}\label{eq:phasedif}
\Delta_{+} - \Delta_{-} = 2m\pi - N\pi,
\end{eqnarray}
We can see that for odd bounces there will be a $\pi$ difference in phase between the reversed orbits caused by boundary, while for even bounces there are no phase differences (ignore the $2\pi$ change). This is in agreement with the analysis of Berry {\it et al.} \cite{Berry:1987}.

\section{Africa A-B Dirac billiard.} \label{Appendix-C}
To confirm our understanding of the mechanism of $\mathcal{T}$-breaking and the magnetic response of relativistic scars, we also analyzed the scars in Africa billiard, a chaotic billiard without geometric symmetry. To obtain the eigenvalues and eigenstates, we did the same calculations (Equations~(\ref{CTM-exp}-\ref{Angle-integral})) as in the heart-shaped billiard.

\begin{figure*}[!htp]
\centering
  \includegraphics[width=0.85\textwidth]{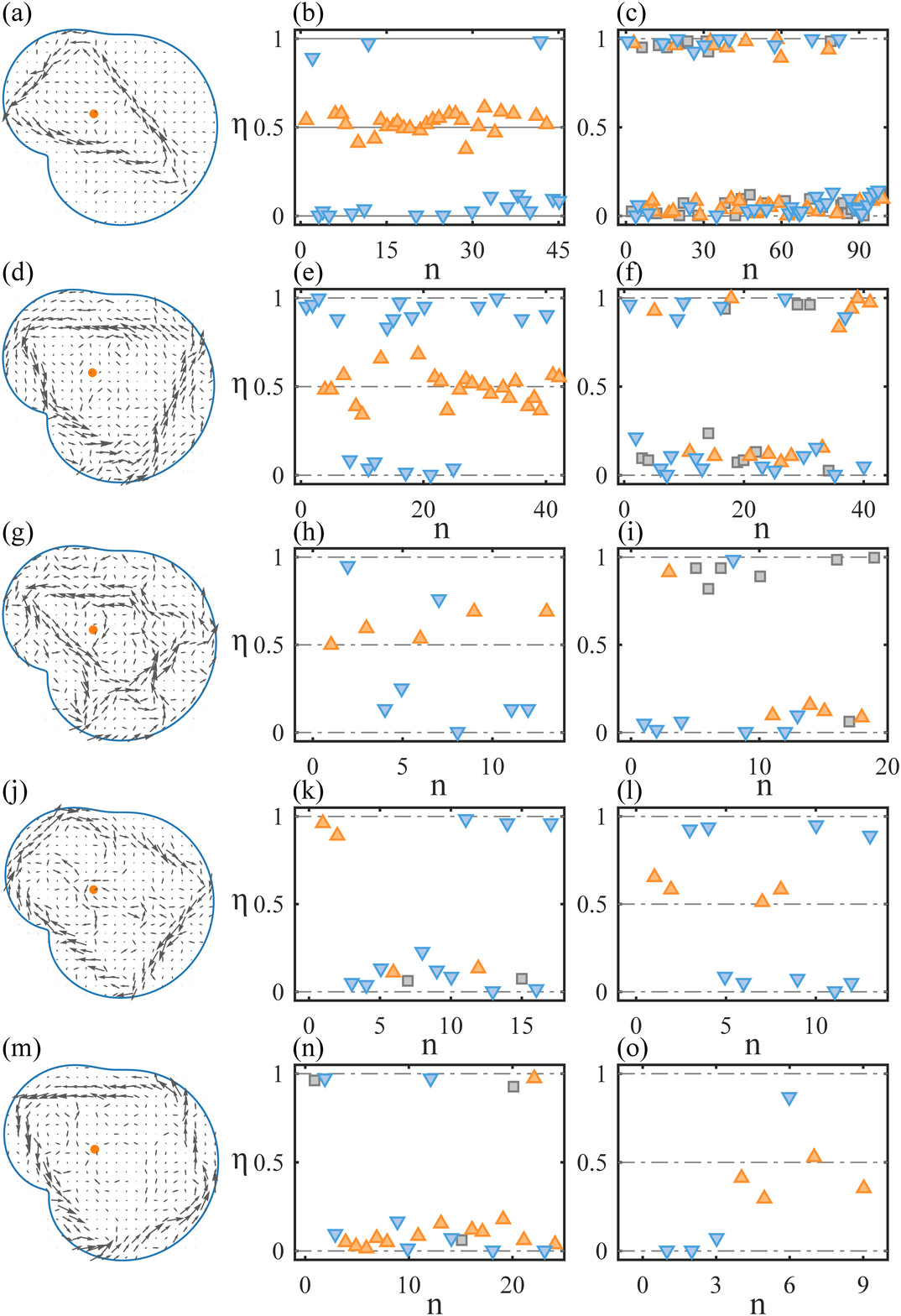}
\caption{The current of scars (a, d, g, j, m), and the corresponding $\eta$ values at $\alpha = 0$ (b, e, h, k, n) and $\alpha = 1/4$ (c, f, i, l, o). The first to the fifth rows are for the period-3-I scar, period-3-II scar, period-5 scar, period-4-I scar and period-4-II scar separately. The orange up-triangles are for scars with counterclockwise flow, the blue down-triangles are for scars with clockwise flow, and the gray squares are for scars whose current orientation is hard to distinguish. The reference state is chosen (arbitrarily) from the scars with clockwise flow.}
\label{fig:Africa-Chiral}
\end{figure*}

Once the eigenstates are obtained, we plot each of them and identify those localized on classical orbits---the scarring states. Then we plot the current flows of the scars and find that for most scars the current has a definitive orientation, as illustrated in figure~{\ref{fig:Africa-Chiral}} (a), (d), (g) for odd period scars and figure~{\ref{fig:Africa-Chiral}} (j), (m) for even period scars. We use $\eta$ (defined in Equation~(\ref{eq:eta})) to characterize the wavevector difference between the repetitive scars on the same orbit. Figure~{\ref{fig:Africa-Chiral}} shows $\eta$ for the scars with counterclockwise flow marked as orange up triangles and those with clockwise flow marked as blue down triangles. First, we consider the zero magnetic flux case. It is found that for even bounce scars, the wavevector difference $\eta$ is always $0$ or $1$, regardless of relative current orientation [figure~{\ref{fig:Africa-Chiral}} (k,n)]; while for odd bounce scars, when two scars have the same current orientation, $\eta = 0$ or $1$, and if two scars have opposite current orientation, then $\eta = 1/2$, as shown in figure~{\ref{fig:Africa-Chiral}} (b,e,h). This current orientation analysis confirms that $\eta = 1/2$ is resulted from the $\pi$ phase difference of the opposite current orientation of odd bounce scars.

Can magnetic flux change the scar chirality in Africa billiard? The answer is yes! By adding a single line of magnetic flux with $\alpha=1/4$ in the origin, the data points of $\eta \sim 0.5$ have been disappeared for odd period scars [figure~{\ref{fig:Africa-Chiral}} (c,f,i)], leading to the lost of chirality and the superficial time-reversal preservation. While for the even (period-4) scars, the data points of $\eta \sim 0.5$ do not present for $\alpha=0$ but emerge for $\alpha=1/4$ [figure~{\ref{fig:Africa-Chiral}} (l,o)]. The interchange of chirality between even and odd period scars indicates that although originated from different mechanism, the boundary-spin interaction induced phase is equivalent to that of magnetic flux. It is noticed that for scars without chiral nature, the two flow orientations are mixed. While for scars with a chiral nature, i.e., odd period orbit scars with $\alpha=0$ and even period scars with $\alpha=1/4$, the scars with different orientations are well separated. One set of the scars attains a $0.5$ value for $\eta$, while the other set attains values of $0$ or $1$.

\begin{figure*}[!htp]
\centering
  \includegraphics[width=1\textwidth]{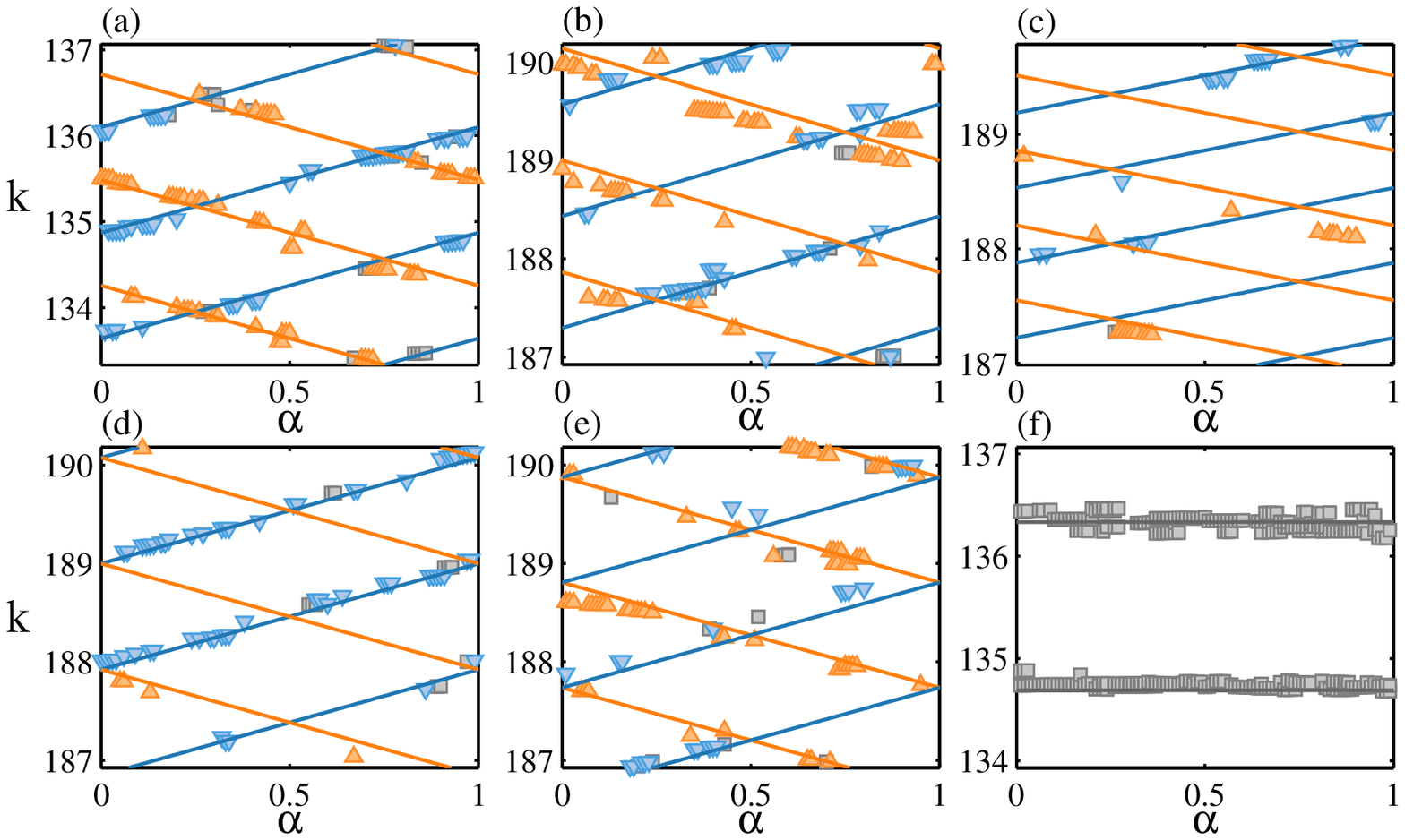}
\caption{The relation between wavevector $k$ and the magnetic flux $\alpha$, for (a) the period-3-I scar shown in figure \ref{fig:Africa-Chiral} (a); (b) the period-3-II scar shown in figure \ref{fig:Africa-Chiral} (d); (c) the period-5 scar shown in figure \ref{fig:Africa-Chiral} (g); (d) the period-4-I scar shown in figure \ref{fig:Africa-Chiral} (j); (e) the period-4-II scar shown in figure \ref{fig:Africa-Chiral} (m); (f) a period-2 scar. The orange up-triangles indicate scars with counterclockwise flow, where $W=1$, and blue down-triangles are the scars with clockwise flow, where $W=-1$. The gray squares are the scars that is difficult to identify the flow orientations. The solid lines are theoretical predictions of Equation (\ref{eq:k-alpha}). The step in the variation of $\alpha$ is 0.01.}
  \label{fig:Africa-k-alpha}
\end{figure*}

In order to have a complete understanding of the chirality and associated phase, we  investigate the magnetic response of scars in a flux interval $0\le \alpha\le1$ (The system is periodic for magnetic flux varying from 0 to 1). From Equation~(\ref{eq:action1}), for a scar with wavevector $k_0$ at $\alpha=0$, as the magnetic flux $\alpha$ is increased, the same scar would appear if the wavevectors approximately follow
\begin{eqnarray}\label{eq:A-k-alpha}
k = k_0 - W \alpha \frac{2\pi}{L},
\end{eqnarray}
where $\beta$ does not appear for it is fixed to a particular value for a certain oriented orbit. We have varied the magnetic flux systematically, and for each case, identify the scars on the same orbit in a certain wavevector (energy) range and identify their flow orientation. The corresponding wavevector versus magnetic flux for the same type scars in figure \ref{fig:Africa-Chiral} are plotted in figure \ref{fig:Africa-k-alpha}. The dashed lines are from Equation~(\ref{eq:A-k-alpha}). We can see that the numerics follow the theory well.

From figure \ref{fig:Africa-k-alpha} it is clear that for the scars on any non-zero closing area orbit, there are actually two sets of scars, one with counterclockwise flow, i.e., $W=1$, where $k$ decreases linearly with increasing $\alpha$; the other with clockwise flow that $W=-1$, where $k$ increases with increasing $\alpha$. Therefore, the two lines cross each other at a certain point, depending on the initial wavevetor value at $\alpha=0$. For the period-3-I and period-3-II scar [figure \ref{fig:Africa-k-alpha} (a,b)], the cross points are $\alpha=0.25$ (corresponding to a $\pi$ phase difference) and $\alpha=0.75$, where the chirality is completely missing. While for the period-4-I and period-4-II scars [figure \ref{fig:Africa-k-alpha} (d,e)], the cross points are at $\alpha=0$ and $\alpha=0.5$. For the period-3 scar, if $\alpha$ is shifted by $0.25$, then the $k$-$\alpha$ relation will behave similarly to that for the period-4 scar, which indicates the accumulated phase difference is $\pi$ for the scars travelling along a complete period with opposite orientation. Here, we should note that for the period-5 scars [figure \ref{fig:Africa-k-alpha} (c)], the $k-\alpha$ relation is similar to that of period-3 scars, as it effectively circulates the flux only one time after a complete orbit. By comparing the different magnetic response of period-5 scar in heart-shaped and Africa billiard, we can see the topological position of the magnetic flux is of vital importance. For period-2 bouncing ball scar [figure \ref{fig:Africa-k-alpha} (f)], as it does not circulate the flux, e.g., $W=0$, thus $k$ does not change with $\alpha$, which agrees with the data.

For Africa billiard, the effect of the boundary induced phase $\beta$ and magnetic phase on scars is similar to that in heart-shaped billiard. This indicates that our understanding of the $\mathcal{T}$-breaking mechanism and the origin of chiral signature in the infinite mass confined billiard is independent of the particular shape of the billiard, although the chirality of the scars can be affected by the number of reflections, the position and magnitude of magnetic flux.

\section{Negative energy, negative potential, mirror symmetry and chiral symmetry}\label{Appendix-D}
Here, we shall provide a comprehensive description of the spin behavior in three cases (and their combinations): negative energy, negative potential and mirror symmetry.

{\bf Negative energy ($-E$), positive potential ($V$) and $V>E>0$.} Considering the action of antiunitary operator $\hat{A} = \hat{\sigma}_x \hat{K}$ on $\hat{H}$
\begin{eqnarray}
\hat{H}' = \hat{A} \hat{H} \hat{A}^{-1} = -\hat{H},
\end{eqnarray}
therefore if $\Psi$ is an eigenstate (especially a scar state) of $\hat{H}$, then it transforms to
\begin{eqnarray*}
\Psi' = \hat{A}
\left( \begin{array}{c}
\psi_{1}\\
\psi_{2}
\end{array} \right)
=\left( \begin{array}{c}
\psi^{*}_{2}\\
\psi^{*}_{1}
\end{array} \right),
\end{eqnarray*}
which is also an eigenstate of $\hat{H}$ with energy $-E$ \cite{Berry:1987}. For the states corresponding to $E$ and $-E$ with the same potential $V$, the probability density distribution is the same: $P = \psi^{*}_{1} \psi_{1} + \psi^{*}_{2} \psi_{2}$. Also, the in-plane probability density current is given by
\begin{eqnarray*}
\bi{u} = v_F\langle \hat{\bi{\sigma}}\rangle = 2v_F[\Re(\psi_{1}^{\ast}(\bi{r}) \psi_{2}(\bi{r})), \Im(\psi_{1}^{\ast}(\bi{r}) \psi_{2}(\bi{r}))],
\end{eqnarray*}
which indicates that the probability density current as well as the in-plane spin behavior at $-E$ is the same as that at $E$ [Equation (\ref{eq:local-current})]. Thus if $\Psi$ is a scar state, the current of scars of these two cases will be the same. Especially, according to Equation~(\ref{eq:SyeigenEV}), if we have $\hat{S}_y\Psi = (\hbar/2)\Psi$ at the boundary ($V \to \infty$), we can also obtain $\hat{S}_y\Psi' = (\hbar/2)\Psi'$. Furthermore, we can get the local expectation value of $\hat{\sigma}_z$. In the positive energy ($E$) case
\begin{eqnarray} \label{eq:sigmazE}
\langle \hat{\sigma}_z \rangle = \Psi^{\dagger}\hat{\sigma}_z \Psi = \psi_{1}^{*}\psi_1 - \psi_{2}^{*}\psi_2,
\end{eqnarray}
while in the negative energy ($-E$) case
\begin{eqnarray} \label{eq:sigmaz-E}
\langle \hat{\sigma}_z \rangle = {\Psi'}^{\dagger}\hat{\sigma}_z \Psi' = \psi_{2}^{*}\psi_2 - \psi_{1}^{*}\psi_1.
\end{eqnarray}
By comparison, we can see that the values of $\langle \hat{\sigma}_z \rangle$ are opposite for $E$ and $-E$ cases. Note that for $E>0$, we can get the explicit local average of $\langle \hat{\sigma}_z \rangle$ at the boundary interface with potential $V$ by using Equation~(\ref{eq:transwave}) and~(\ref{eq:Transmission}),
\begin{eqnarray}
\langle \hat{\sigma}_z \rangle = \Psi^{\dagger}_{2} \hat{\sigma}_z \Psi_{2} = 4\cos^{2}\gamma \frac{Eq - VK}{(V - E)(q + K)}.
\end{eqnarray}
We can prove that $\langle \hat{\sigma}_z \rangle \ge 0$. Especially, when $V \to +\infty$, $\langle \hat{\sigma}_z \rangle = 0$. Similarly, for $E<0$, we have $\langle \hat{\sigma}_z \rangle \le 0$ and $\langle \hat{\sigma}_z \rangle = 0$ when $V \to +\infty$.

\begin{figure}[h]
\centering
  \includegraphics[width=0.9\textwidth]{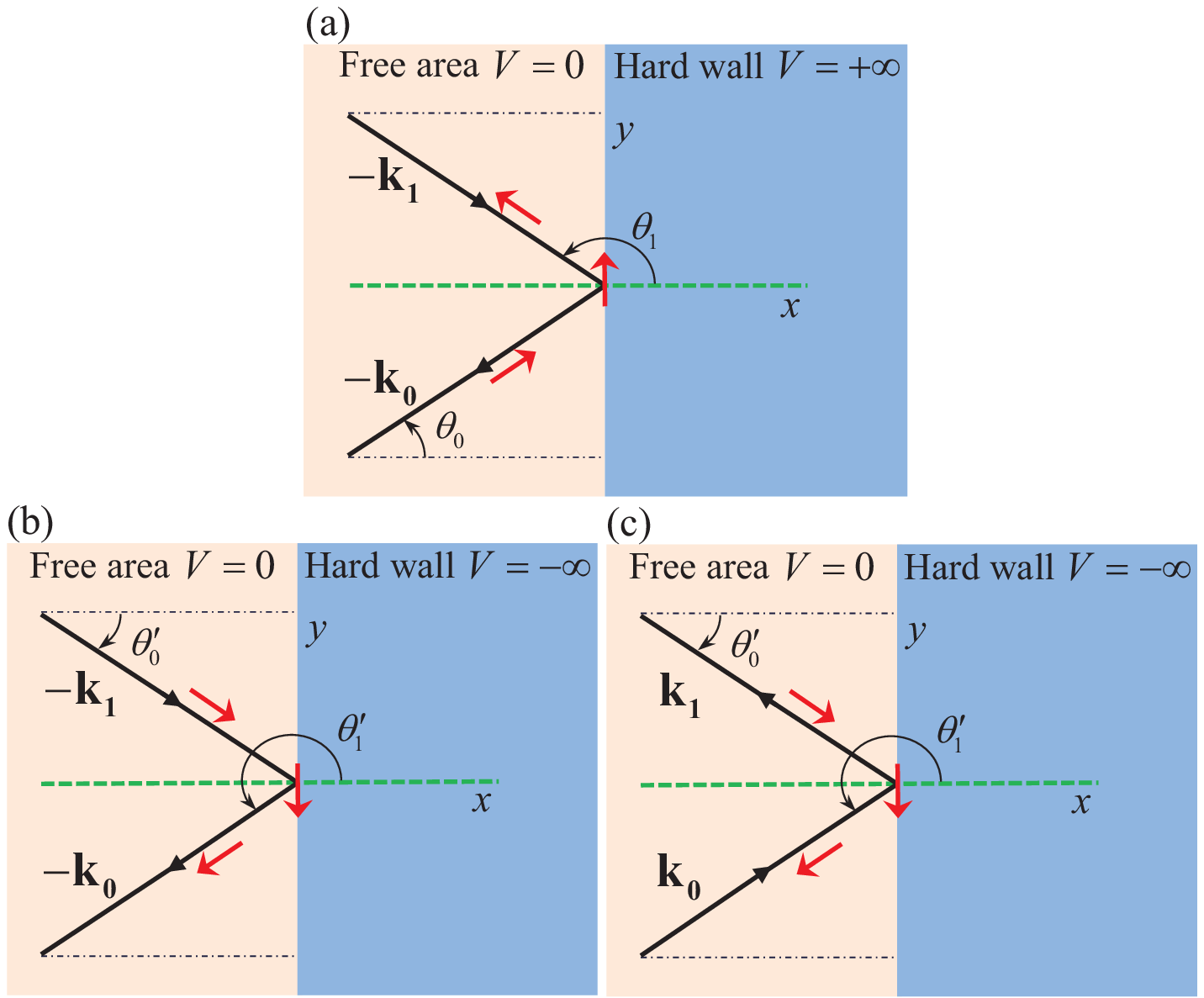}
\caption{Incident and reflected plane waves (black arrow) and the spin (red arrow) corresponding to the superposition of waves at the boundary. (a) $-E$, $V$, and $V\to+\infty$; (b) $E$, $-V$, and $V\to+\infty$; (c) $-E$, $-V$, and $V\to+\infty$.}
  \label{fig:planewave-EV}
\end{figure}

We now investigate the spin behavior at the boundary and, most importantly, compare the accumulated phase difference of the scar orbits with respect to $E$ and $-E$ by employing the plane wave model as proposed in~Equations ({\ref{eq:planewave}}-~{\ref{eq:R-coe}}). Note that the helicity is $\bi{\sigma}\cdot \bi{p}/|\bi{p}| = -1$ at $-E$ compared with the positive energy case where $\bi{\sigma}\cdot \bi{p}/|\bi{p}| = 1$ for the free particle, which means although the current orientation is the same for $E$ and $-E$, the momentum of the free particle is in reversed current direction in $-E$ case. The wavefunction in the plain area at $-E$ is
\begin{eqnarray} \label{eq:planewave-E}
\Psi_1 = & \frac{1}{\sqrt{2}}\Bigg[
\left( \begin{array}{c}
\exp{\{-\frac{1}{2}i(\theta_0)\}}\\
\exp{\{\frac{1}{2}i(\theta_0)\}}
\end{array} \right)\exp{\{-i\bi{k}_0\cdot \bi{r}\}}
\nonumber \\
& + R
\left( \begin{array}{c}
\exp{\{-\frac{1}{2}i(\theta_1)\}}\\
\exp{\{\frac{1}{2}i(\theta_1)\}}
\end{array} \right)
\exp{\{-i\bi{k}_1\cdot \bi{r}\}}\Bigg],
\end{eqnarray}
as illustrated in figure~\ref{fig:planewave-EV}(a), where we adopt $\theta_0$ and $\theta_1$ as the spin direction of the free particle, $\theta_0 + \pi$ and $\theta_1 + \pi$ as its wavevector direction. The transmitted wave in the potential area is
\begin{eqnarray}
\Psi_2 = \frac{T}{\sqrt{2}}
\left( \begin{array}{c}
-i\lambda_1\\
\lambda_2
\end{array} \right)
e^{-qx}e^{iKy},
\end{eqnarray}
where the wavevector $-\bi{k}_0 = (-k\cos\theta_0, -k\sin\theta_0)$ and $-\bi{k}_1 = (-k\cos\theta_1, -k\sin\theta_1)$, $K = -k\sin\theta_0$ and $q = \sqrt{\frac{V^2 - E^2}{\hbar^2v_F^2} + K^2}$; $E = \hbar v_F k$, and $\lambda_1 = \sqrt{\frac{(V - E)(q - K)}{Vq + EK}}$; $\lambda_2 = \sqrt{\frac{(V + E)(q + K)}{Vq + EK}}$.
Using the convention~(\ref{eq:specularity}) and matching the wavefunctions $\Psi_1$ and $\Psi_2$ at the boundary, we can obtain the formula of $R$ and $T$. Especially in $V\to +\infty$ limit, we can get $R = 1$. Now, we can verify the spin orientation at the boundary using the convention Equation~(\ref{eq:specularity}) and $R = 1$, and we have
\begin{eqnarray}
\hat{S}_y \Psi_1 = \frac{\hbar}{2} \Psi_1.
\end{eqnarray}
Thus the spin points to the positive $y$-axis direction at the boundary in the $-E$ case with $V \to \infty$.
Furthermore, we can get the phase change for the scar state with counterclockwise current
\begin{eqnarray}\label{eq:positivephase-E}
\Delta_{+} = \frac{1}{2}(\theta_{n} - \theta_{0}) = m\pi,
\end{eqnarray}
where $m$ is an integer. By comparing Equation~({\ref{eq:positivephase-E}}) with Equation~({\ref{eq:positivephase}}), we can see that the accumulated phase of the two orbits with the same current orientation corresponding to $E$ and $-E$ is the same. For the reversed orbit with clockwise flow, the incident and reflected angles are defined as $\theta'_0$ and $\theta'_1$, which are the same as that defined in figure~\ref{fig:planewave}. By using the relations Equation~({\ref{eq:localangle1}}) and Equation~({\ref{eq:localangle2}}), we can get the phase change
\begin{eqnarray}\label{eq:negativephase-E}
\Delta_{-} = \frac{1}{2}(\theta'_{n} - \theta'_{0})= -m\pi + N\pi,
\end{eqnarray}
where $N$ is the number of reflections along the orbit. This is the same as Equation~(\ref{eq:negativephase}). So the accumulated phase difference between the reversed orbits caused by the boundary at $-E$ case is
\begin{eqnarray}\label{eq:phasedifC}
\Delta_{+} - \Delta_{-} = 2m\pi - N\pi.
\end{eqnarray}
For odd orbit ($N$ is odd), there is an additional $\pi$ difference between the counterclockwise state and the clockwise state, thus the chiral scars still exist.

{\bf Positive energy ($E$), negative potential ($-V$) and $V>E>0$.} The time reversal operator is defined as $\hat{T} = i\hat{\sigma}_y \hat{K}$. Under the action of $\hat{T}$, $\hat{H}$ transforms to
\begin{eqnarray} \label{negative-V}
\hat{H}' = \hat{T} \hat{H} \hat{T}^{-1} = v_F \hat{\bi{\sigma}} \cdot \hat{\bi{p}} - V(\bi{r}) \hat{\sigma}_z,
\end{eqnarray}
and the eigenstate $\Psi$ of $\hat{H}$ transforms to
\begin{eqnarray}
\Psi' = \hat{T}
\left( \begin{array}{c}
\psi_{1}\\
\psi_{2}
\end{array} \right)
=\left( \begin{array}{c}
\psi^{*}_{2}\\
-\psi^{*}_{1}
\end{array} \right),
\end{eqnarray}
where $\hat{H}'\Psi'=E\Psi'$. We can see the probability distribution is the same for $\Psi$ and $\Psi'$: $P = \psi^{*}_{1} \psi_{1} + \psi^{*}_{2} \psi_{2}$, while the current orientation is opposite,
\begin{eqnarray}
\bi{u}' = -2v_F[\Re(\psi_{1}^{\ast}(\bi{r}) \psi_{2}(\bi{r})), \Im(\psi_{1}^{\ast}(\bi{r}) \psi_{2}(\bi{r}))] = -\bi{u}.
\end{eqnarray}
We can also obtain the local expectation value of $\langle \hat{\sigma}_z \rangle$
\begin{eqnarray}
\langle \hat{\sigma}_z \rangle = {\Psi'}^{\dagger}\hat{\sigma}_z \Psi' = \psi_{2}^{*}\psi_2 - \psi_{1}^{*}\psi_1,
\end{eqnarray}
which is the same as that in ($-E$, $V$) case (Equation~(\ref{eq:sigmaz-E})) and opposite to ($E$, $V$) case (Equation~(\ref{eq:sigmazE})).

Now, let us examine the spin orientation at the boundary in the framework of plane wave and then calculate the accumulated phase along the periodic orbit in the negative potential billiard. The wavefunction in the free area can be written as
\begin{eqnarray} \label{eq:planewave-V}
\Psi_1 = & \frac{1}{\sqrt{2}}\Bigg[
\left( \begin{array}{c}
\exp{\{-\frac{1}{2}i(\theta'_0)\}}\\
\exp{\{\frac{1}{2}i(\theta'_0)\}}
\end{array} \right)\exp{\{-i\bi{k}_1\cdot \bi{r}\}}
\nonumber \\
& + R
\left( \begin{array}{c}
\exp{\{-\frac{1}{2}i(\theta'_1)\}}\\
\exp{\{\frac{1}{2}i(\theta'_1)\}}
\end{array} \right)
\exp{\{-i\bi{k}_0\cdot \bi{r}\}}\Bigg],
\end{eqnarray}
as illustrated in figure \ref{fig:planewave-EV}(b). And the transmitted wave in the potential area has the form
\begin{eqnarray}\label{eq:transwave-V}
\Psi_2 = \frac{T}{\sqrt{2}}
\left( \begin{array}{c}
i\lambda_1\\
\lambda_2
\end{array} \right)
e^{-qx}e^{iKy},
\end{eqnarray}
where the incident wave vector $-\bi{k}_1 = (k\cos\theta'_0, k\sin\theta'_0)$ and the reflected wave vector $-\bi{k}_0 = (k\cos\theta'_1, k\sin\theta'_1)$, $K = k\sin\theta'_0$ and $q = \sqrt{\frac{V^2 - E^2}{\hbar^2v_F^2} + K^2}$; $E = \hbar v_F k$, and $\lambda_1 = \sqrt{\frac{(V - E)(q - K)}{Vq + EK}}$; $\lambda_2 = \sqrt{\frac{(V + E)(q + K)}{Vq + EK}}$. Matching the waves at the boundary and using the specularity~(\ref{eq:specularity}), we can obtain the reflection and transmission coefficients.
Especially when we take $V\to - \infty$, we can get $R=-1$. From~\ref{Appendix-B}, we know that a $\pi$ phase in the reflection wave can reverse the spin orientation. So, at the boundary it is still an eigenfunction of $\hat{S}_y$ but with an eigenvalue of $-\hbar/2$,
\begin{eqnarray*}
\hat{S}_y\Psi = -\frac{\hbar}{2}\Psi.
\end{eqnarray*}
regardless of the incident angle.
According to Equation~(\ref{eq:negativephase}), the whole phase change caused by boundary along the clockwise orientation is
\begin{eqnarray}\label{eq:negativephase-V}
\Delta_{-} = \frac{1}{2}(\theta^{'}_{n} - \theta^{'}_{0}) - N\pi= -m\pi,
\end{eqnarray}
whereas the phase change along the counterclockwise direction is
\begin{eqnarray}\label{eq:positivephase-V}
\Delta_{+} = \frac{1}{2}(\theta_{n} - \theta_{0}) - N\pi= m\pi - N\pi.
\end{eqnarray}
Because $\Delta_{+} - \Delta_{-} = 2m\pi - N\pi$, the chirality of odd orbits still exists in the negative potential case. In addition, by comparing Equation~({\ref{eq:negativephase-V}}) with Equation~({\ref{eq:positivephase}}), we can see that the difference of the accumulated phase of these two orbits with opposite current orientation corresponding to ($E$, $V$) and ($E$, $-V$) is an integer multiple of $2\pi$.

{\bf Negative energy ($-E$), negative potential ($-V$) and $V>E>0$.} Applying the unitary operator $\hat{U} = i\hat{\sigma}_y \hat{\sigma}_x$ combined of two antiunitary operators
$\hat{\sigma}_x \hat{K}$ and $i\hat{\sigma}_y \hat{K}$ to act on the Hamiltonian, we can get
\begin{eqnarray}
\hat{H}'' = \hat{A}\hat{H}\hat{A}^{-1} = -(v_F \hat{\bi{\sigma}} \cdot \hat{\bi{p}} - V\hat{\sigma}_z),
\end{eqnarray}
and the eigenstate $\Psi$ of $\hat{H}$ changes into
\begin{eqnarray}
\Psi' = \hat{U}
\left( \begin{array}{c}
\psi_{1}\\
\psi_{2}
\end{array} \right)
=\left( \begin{array}{c}
\psi_{1}\\
-\psi_{2}
\end{array} \right).
\end{eqnarray}
Therefore, $\Psi'$ is the eigenstate of $\hat{H}' = v_F \hat{\bi{\sigma}} \cdot \hat{\bi{p}} - V$ with negative energy $-E$. The probability is still the same as that in ($E$, $V$) case, i.e. $P = \psi^{\ast}_1\psi_1 + \psi^{\ast}_2\psi_2$, while the current orientation is opposite,
\begin{eqnarray}
\bi{u}' = -2v_F[\Re(\psi_{1}^{\ast}(\bi{r}) \psi_{2}(\bi{r})), \Im(\psi_{1}^{\ast}(\bi{r}) \psi_{2}(\bi{r}))] = -\bi{u}.
\end{eqnarray}
Also, the local expectation value of $\hat{\sigma}_z$ is
\begin{eqnarray}
\langle \hat{\sigma}_z \rangle = {\Psi'}^{\dagger}\hat{\sigma}_z \Psi' = \psi^{\ast}_1 \psi_1 - \psi^{\ast}_2\psi_2,
\end{eqnarray}
which indicates that $\langle \hat{\sigma}_z \rangle$ is the same as that in ($E$, $V$) case.

To confirm the spin behavior at the boundary and obtain global phase change of spin along a complete periodic orbit, we use the plane wave model with the wave in the free area written as
\begin{eqnarray} \label{eq:planewave-EV}
\Psi_1 = & \frac{1}{\sqrt{2}}\Bigg[
\left( \begin{array}{c}
\exp{\{-\frac{1}{2}i(\theta'_0)\}}\\
\exp{\{\frac{1}{2}i(\theta'_0)\}}
\end{array} \right)\exp{\{i\bi{k}_1\cdot \bi{r}\}}
\nonumber \\
& + R
\left( \begin{array}{c}
\exp{\{-\frac{1}{2}i(\theta'_1)\}}\\
\exp{\{\frac{1}{2}i(\theta'_1)\}}
\end{array} \right)
\exp{\{i\bi{k}_0\cdot \bi{r}\}}\Bigg].
\end{eqnarray}
The schematic diagram is shown in figure~\ref{fig:planewave-EV}(b) and the transmitted wave in the potential area is in the form
\begin{eqnarray}\label{eq:transwave-EV}
\Psi_2 = \frac{T}{\sqrt{2}}
\left( \begin{array}{c}
i\lambda_1\\
\lambda_2
\end{array} \right)
e^{-qx}e^{iKy},
\end{eqnarray}
where the wave vector $\bi{k}_0 = (-k\cos\theta'_0, -k\sin\theta'_0)$, $\bi{k}_1 = (-k\cos\theta'_1, -k\sin\theta'_1)$, $K = -k\sin\theta'_0$ and $q = \sqrt{\frac{V^2 - E^2}{\hbar^2v_F^2} + K^2}$; $E = \hbar v_F k$, and $\lambda_1 = \sqrt{\frac{(V + E)(q - K)}{Vq - EK}}$; $\lambda_2 = \sqrt{\frac{(V - E)(q + K)}{Vq - EK}}$. Matching the wavefunctions at the boundary and using the specularity~(\ref{eq:specularity}), we can get $R=-1$ when $V \to \infty$. So the whole phase change caused by boundary along the clockwise orientation is
\begin{eqnarray}\label{eq:negativephase-EV}
\Delta_{-} = \frac{1}{2}(\theta^{'}_{n} - \theta^{'}_{0}) - N\pi= -m\pi,
\end{eqnarray}
which is the same as Equation~(\ref{eq:negativephase-V}) in ($E$, $-V$) case. The accumulated phase along the counterclockwise direction is
\begin{eqnarray}\label{eq:positivephase-EV}
\Delta_{+} = \frac{1}{2}(\theta_{n} - \theta_{0}) - N\pi= m\pi - N\pi,
\end{eqnarray}
which is the same as Equation~(\ref{eq:positivephase-V}). The chirality of odd orbits still exists in the ($-E$, $-V$) case. By comparing Equation~({\ref{eq:negativephase-EV}}) with Equation~({\ref{eq:positivephase}}), we can see that the difference of the accumulated phase of the two orbits with the opposite current orientation corresponding to ($-E$, $-V$) and ($E$, $V$) is still an integer multiple of $2\pi$.

\begin{figure}[h]
\centering
  \includegraphics[width=0.75\textwidth]{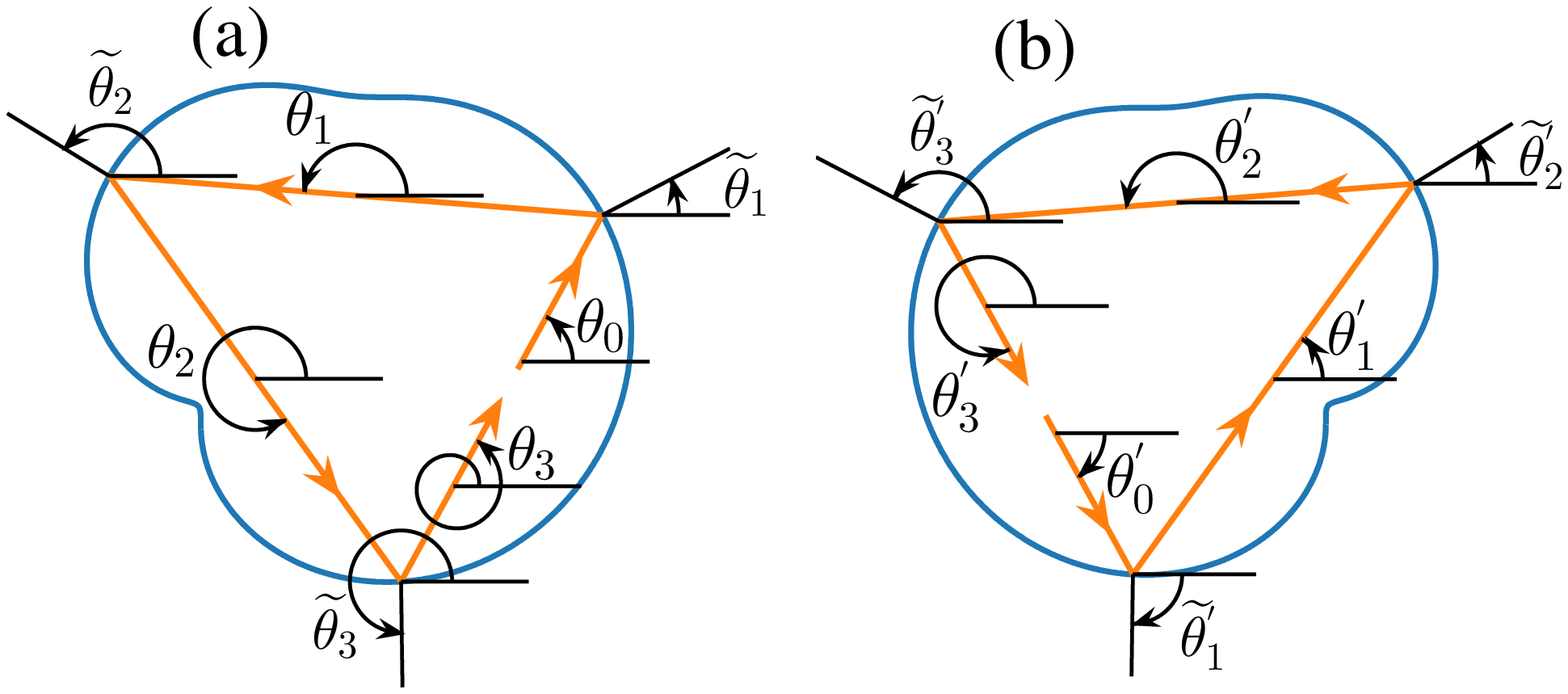}
\caption{The Africa-shape billiard with counterclockwise flow. (a) The original scar orbit; (b) The state under mirror reflection $\hat{A} = \hat{R}_{x}\hat{K}$.}
\label{fig:Africa-parity}
\end{figure}

{\bf Mirror reflection}. The potential $V(x, y)$ changes to $V(-x, y)$ under the action of antiunitary operator $\hat{A} = \hat{R}_{x}\hat{K}$, i.e.,
\begin{eqnarray}
\hat{H}' = \hat{A} \hat{H} \hat{A}^{-1} = v_F \hat{\bi{\sigma}}\cdot \hat{\bi{p}} + V(-x, y),
\end{eqnarray}
as illustrated in figure~(\ref{fig:Africa-parity}). The eigenstate $\Psi$ of $\hat{H}$ transforms to
\begin{eqnarray}
\Psi' = \hat{R}_x \hat{K}
\left( \begin{array}{c}
\psi_{1}(x,y)\\
\psi_{2}(x,y)
\end{array} \right)
=\left( \begin{array}{c}
\psi^{*}_{1}(-x,y)\\
\psi^{*}_{2}(-x,y)
\end{array} \right)
\end{eqnarray}
where $\hat{H}'\Psi'=E\Psi'$. The probability distribution is $P = \psi^{\ast}_1(-x, y)\psi_1(-x,y) + \psi^{\ast}_2(-x,y)\psi_2(-x,y)$, which is also symmetric about $y$ axis. The local current can be written as
\begin{eqnarray}
\bi{u}'
& = 2v_F\big[\Re\big(\psi_{1}(-x,y) \psi_{2}^{\ast}(-x,y)\big), \Im\big(\psi_{1}(-x, y) \psi_{2}^{\ast}(-x, y)\big)\big] \nonumber\\
& = 2v_F\big[\Re\big(\psi_{1}^{\ast}(-x, y) \psi_{2}(-x, y)\big), -\Im\big(\psi_{1}^{\ast}(-x,y) \psi_{2}(-x,y))\big].
\end{eqnarray}
Thus $u'_x(x,y) = u_x(-x,y)$, and $u'_y(x,y) = -u_y(-x,y)$. 
This indicates that the current of the scars with the same energy of these two systems is in the same winding orientation, as shown in figure~{\ref{fig:Africa-parity}}. The accumulated phase difference around a complete periodic orbit of these two systems with the same winding direction can be obtained as follows (the schematic diagram can be seen in figure~{\ref{fig:Africa-parity}}). First, for the odd orbits of the system with Hamiltonian $\hat{H}$ as shown in figure~{\ref{fig:Africa-parity}}(a), the accumulated phase~{\cite{Berry:1987}} is
\begin{eqnarray}
\Delta_o =\frac{1}{2}(\theta_N - \theta_0) = \frac{1}{2} \bigg(\pi - 2\theta_0 + 2\Big(\sum^{M}_{j=1}{\widetilde{\theta}_{2j-1}} - \sum^{M-1}_{j=1}{\widetilde{\theta}_{2j}}\Big)\bigg).
\end{eqnarray}
While for the system $\hat{H}'$ as illustrated in figure~{\ref{fig:Africa-parity}}(b), the accumulated phase along the complete orbit is
\begin{eqnarray}
\Delta'_o = \frac{1}{2}(\theta'_N - \theta'_0) = \frac{1}{2} \bigg(-\pi + 2\theta_0 - 2\Big(\sum^{M}_{j=1}{\widetilde{\theta}_{2j-1}} - \sum^{M-1}_{j=1}{\widetilde{\theta}_{2j}}\Big) + 4m\pi\bigg),
\end{eqnarray}
where we have used the angle relations $\theta'_0 = -\theta_0$ and $\widetilde{\theta}_{2j-1} = \pi - \widetilde{\theta}_{2(M-j+1)-1} + 2n\pi$. Closure means $\Delta_o = K\pi$ ($K$ is integer), as a result of which we can obtain the accumulated phase difference of these two orbits
\begin{eqnarray}
\Delta_o - \Delta'_o = 2K\pi - 2m\pi,
\end{eqnarray}
which illustrates that the accumulated phase difference of these two orbits are multiple integers of $2\pi$.
For the even orbits, the accumulated phase for the system with Hamiltonian $\hat{H}$ is
\begin{eqnarray}
\Delta_e =\frac{1}{2}(\theta_N - \theta_0) = \Big(\sum^{M}_{j=1}{\widetilde{\theta}_{2j} - \widetilde{\theta}_{2j-1}}\Big),
\end{eqnarray}
while for the system $\hat{H}'$, by using $\widetilde{\theta}_{2j-1} = -\widetilde{\theta}_{2(M-j+1)}+\pi + 2n\pi$, we can get
\begin{eqnarray}
\Delta'_e =\frac{1}{2}(\theta'_N - \theta'_0) = \Big(\sum^{M}_{j=1}{\widetilde{\theta}_{2j} - \widetilde{\theta}_{2j-1}}  + 2m'\pi\Big).
\end{eqnarray}
The accumulated phase difference is still an integer multiple of $2\pi$. However, if in figure~{\ref{fig:Africa-parity}}(b) the current flow has an opposite orientation and it is an odd orbit scar, then it will have an additional $\pi$ phase difference compared with the scar in figure~{\ref{fig:Africa-parity}}(a). While for even orbits this $\pi$ phase difference does not appear.

\begin{figure}[h]
\centering
  \includegraphics[width=0.75\textwidth]{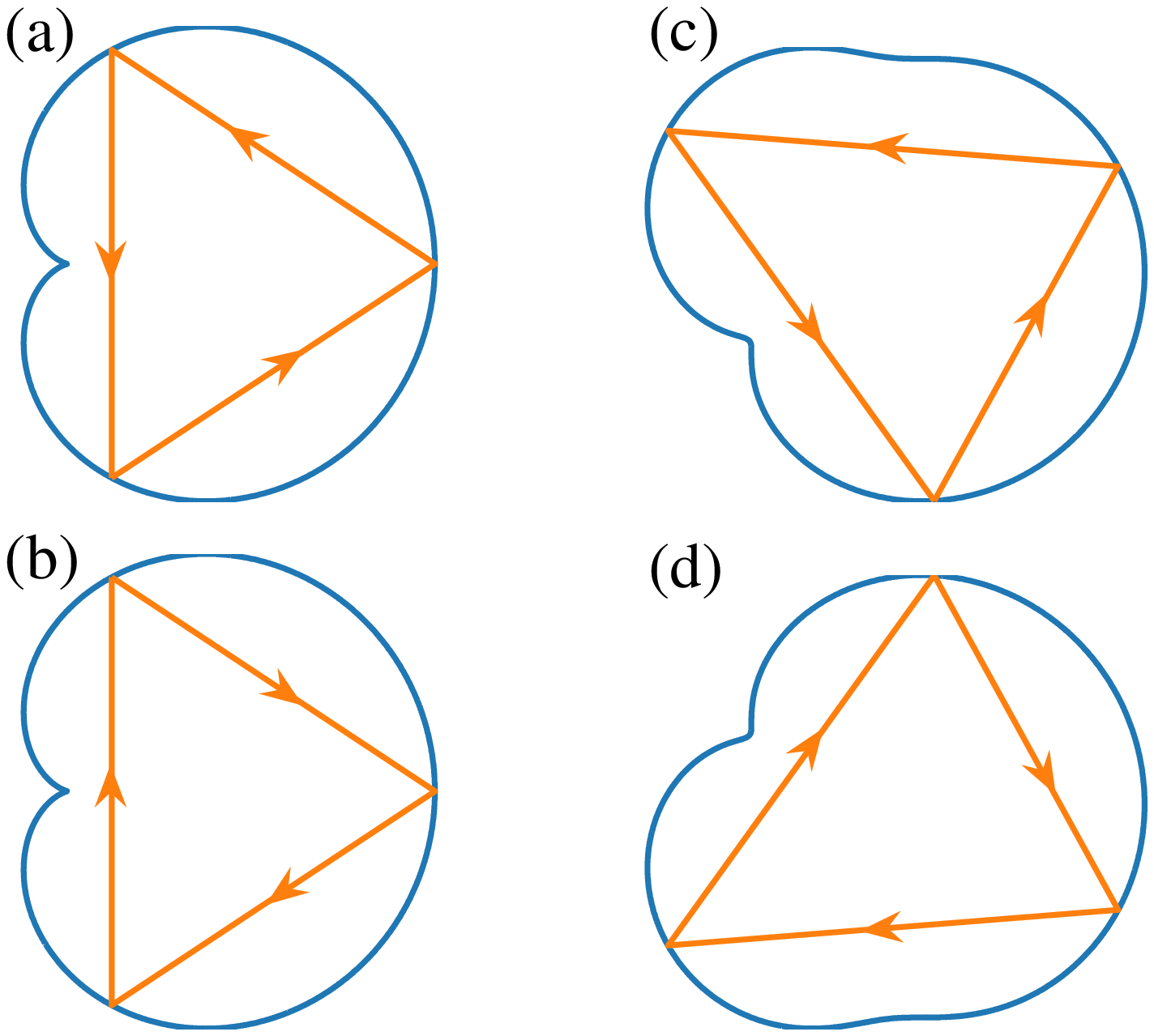}
\caption{The heart-shaped billiard [(a), (b)] and Africa billiard [(c), (d)] with certain flow orientation. (a), (c): Scar orbit of the original billiard at ($E,V(x,y)$); (b), (d): The scar orbit under parity transformation ($\hat{P} = \hat{R}_{y}\hat{\sigma}_x$) at ($E, -V(x,-y)$).}
  \label{fig:HA-parity}
\end{figure}

{\bf Parity operation}. Here, we consider the parity operation with respect to the $x$ axis. The parity operator is $\hat{P}=\hat{R}_y \hat{\sigma}_x$. Under its action, the Hamiltonian $\hat{H}$ transforms to
\begin{eqnarray}
\hat{H}' = \hat{P} \hat{H} \hat{P}^{-1} = v_F \hat{\bi{\sigma}}\cdot \hat{\bi{p}} - V(x, -y),
\end{eqnarray}
and the eigenstate $\Psi$ of $\hat{H}$ transforms to
\begin{eqnarray}
\Psi' = \hat{P}
\left( \begin{array}{c}
\psi_{1}(x,y)\\
\psi_{2}(x,y)
\end{array} \right)
=\left( \begin{array}{c}
\psi^{*}_{2}(x,-y)\\
\psi^{*}_{1}(x,-y)
\end{array} \right),
\end{eqnarray}
where $\hat{H}'\Psi'=E\Psi'$. It is noticed that after the parity operation, beside the mirror reflection with respect to the $x$ axis, the confinement potential changes sign, indicating parity symmetry is broken. Effectively, the parity operation is equivalent to the mirror reflection together with the time-reversal operation, which changes the sign of $V$. The probability distribution is $P = \psi^{\ast}_1(x,-y)\psi_1(x,-y) + \psi^{\ast}_2(x,-y)\psi_2(x,-y)$, which is also symmetric about $x$ axis. The local current can be written as
\begin{eqnarray}
\bi{u}'
& = 2v_F\big[\Re\big(\psi_{1}(x,-y) \psi_{2}^{\ast}(x,-y)\big), \Im\big(\psi_{1}(x,-y) \psi_{2}^{\ast}(x,-y)\big)\big] \nonumber\\
& = 2v_F\big[\Re\big(\psi_{1}^{\ast}(x,-y) \psi_{2}(x,-y)\big), -\Im\big(\psi_{1}^{\ast}(x,-y) \psi_{2}(x,-y))\big].
\end{eqnarray}
Thus $u'_x(x,y) = u_x(x,-y)$, and $u'_y = -u_y(x,-y)$. The schematic diagram of the scar current is shown in figure~{\ref{fig:HA-parity}}. For the heart-shaped billiard, $V(x,y) = V(x,-y)$, so the Hamiltonian $\hat{H}'$ is the same as that under $\mathcal{T}$ operation (Equation~(\ref{negative-V})). This indicates that $\mathcal{T}$ and parity operation have the same effect for the heart-shaped billiard, and the system is invariant under the combination of $\mathcal{P}$ and $\mathcal{T}$ operation ($\hat{A} = \hat{R}_y \hat{\sigma}_x\cdot i\hat{\sigma}_y \hat{K}=-\hat{R}_y \hat{\sigma}_z \hat{K}$). For Africa billiard, the scar current orientation of $\hat{H}'$ is opposite to the original system. If we rotate the Africa billiard in figure~{\ref{fig:HA-parity} (d)} by $\pi$, we can get the same geometric shape as the billiard in figure~{\ref{fig:Africa-parity} (b)}. The difference is the sign of the potential, thus the current direction is opposite for these two cases. Note that we can also use the parity operator $\hat{P}' = \hat{R}_x \hat{\sigma}_y$, which gives the parity operation with respect to $y$ axis. The action of $\hat{P}'$ equals to the combination of $\hat{P}$ and a rotation by $\pi$. The mirror operator in fact is the combination of parity and time-reversal operators, i.e., $\hat{A} = \hat{R}_x\hat{K} = \hat{R}_x \hat{\sigma}_y \cdot \hat{\sigma}_y {K}$.

Furthermore, we have considered all the combinations of $\pm E$, $\pm V$, with or without $\mathcal{M}$, the results are summarized in table~{\ref{tab:EVP}}.
\begin{table}
\centering
\caption{The spin properties of different combinations of three operations. $E$ is the energy of the system, $V$ is the potential and $\mathcal{M}$ is the mirror reflection with regard to $y$-axis, $\mathcal{I}$ is without the $\mathcal{M}$ operation. $R$ is the reflection coefficient in the planewave model with convention Equation~({\ref{eq:specularity}}); helicity is defined as $\hat{\bi{\sigma}}\cdot \hat{\bi{p}}/|\bi{p}|$ in the free billiard domain ($V=0$), if helicity is $\pm 1$, the direction of the wavevector is the same (opposite) with the current orientation. Scar current means the current orientation in the billiard domain with $\pm$ indicates the same (opposite) orientation as the ($E$, $V$, $\mathcal{I}$) case. Spin orientation is the direction of spin at the boundary interface and $+$ represents positive $y$-direction (counterclockwise orientation with respect to outer normal vector). $\langle \hat{\sigma}_z \rangle$ represents the local average of spin in the $z$ direction at the boundary interface when the potential is finite, and $\pm$ is for positive (negative) $z$ axis. When $V\to \infty$, $\langle \hat{\sigma}_z \rangle = 0$.}
\label{tab:EVP}
\renewcommand{\arraystretch}{1.3}
\begin{tabular*}{0.75\textwidth}{@{}c@{\extracolsep{\fill}}cccccccc}
\br
 &$E$  &$E$   &$E$  &$E$  &$-E$  &$-E$  &$-E$  &$-E$\\ 
 &$V$  &$-V$  &$V$  &$-V$ &$V$   &$-V$  &$V$   &$-V$\\ 
 &$\mathcal{I}$ &$\mathcal{I}$  &$\mathcal{M}$  &$\mathcal{M}$  &$\mathcal{I}$  &$\mathcal{I}$  &$\mathcal{M}$   &$\mathcal{M}$\\
 \mr
Reflection (R)                  &1  &-1 &1  &-1 &1 &-1  &1 &-1 \\
Helicity                        &1  &1  &1  &1  &-1  &-1  &-1  &-1 \\
Scar current               &+  &-  &+  &-  &+  &-  &+  &- \\
Spin orientation                &+  &-  &+  &-  &+  &-  &+  &- \\
$\langle \hat{\sigma}_z \rangle$        &+  &-  &+  &-  &-  &+  &-  &+ \\
\br
\end{tabular*}
\end{table}

%
%
\section*{References}
\bibliography{iopart-num}

\begin{thebibliography}{10}
\bibitem{ref:TB}
Although in a more rigorous way the magnetic field also changes sign under
  $\mathcal{T}$ operation, here we assume that the magnetic field is an
  external field and is fixed, and the $\mathcal{T}$ operator only applies on
  the motion of the billiard.
\end{thebibliography}

\end{document}